\documentclass[10pt]{IEEEtran}
\def\figureref#1{Fig.~\ref{#1}}
\usepackage[dvips]{graphicx}
\usepackage{epsf, rotate, url} 
\usepackage{amsmath,subfigure,url}
\usepackage{amsthm} 
\usepackage{amsfonts}
\usepackage{amssymb}
\usepackage{newlfont}
\usepackage{algorithmic, algorithm}
\usepackage{authblk}
\usepackage{xcolor}
\theoremstyle{plain}

\theoremstyle{remark}

\long\def\comment#1{}
\long\def\commentHet#1{}

\long\def\commentHQ#1{\textbf{HQ-TextCheck} }
\long\def\commentHQok#1{#1}
\long\def\commentHQdelete#1{}
\def\figureref#1{Fig.~\ref{#1}}
\def\eqref#1{(\ref{#1})}

\IEEEoverridecommandlockouts
\title{Energy-Aware Aggregation of Dynamic Temporal Workload in Data Centers}
\author{Haiyang Qian$^1$, Fu Li$^2$, Ravishankar Ravindran$^3$, and Deep Medhi$^{1,4}$
\thanks{$^1$University of Missouri--Kansas City, $^2$University of Wisconsin--Madison;
 $^3$Huawei Innovation Center; $^4$Indian Institute of Technology-Guwahati}
\thanks{$^1$Supported in part by NSF Grant No. CNS-0916505. Haiyang Qian is now with China Mobile USA Research Center}}

\comment{\usepackage[pscoord]{eso-pic}
\newcommand{\placetextbox}[3]{
  \setbox0=\hbox{#3}
  \AddToShipoutPictureFG*{
    \put(\LenToUnit{#1\paperwidth},\LenToUnit{#2\paperheight}){\vtop{{\null}\makebox[0pt][c]{#3}}}%
  }%
}%
\placetextbox{.2}{0.055}{978-3-901882-48-7~\copyright~2012 IFIP}%
}

\begin{document}
\maketitle
\begin{abstract}
Data center providers seek to minimize their total cost of
ownership (TCO), while  power consumption has become a social
concern. We present formulations to minimize server energy
consumption and server cost under three different data center
server setups (homogeneous, heterogeneous, and hybrid
hetero-homogeneous clusters) with dynamic temporal workload.
Our studies show that the homogeneous model significantly
differs from the heterogeneous model in computational time (by
an order of magnitude). To be able to compute optimal
configurations in near real-time for large scale data centers,
we propose two modes, aggregation by maximum and aggregation by
mean. \comment{heuristics: 1) aggregation by maximum mode and
2) aggregation by mean mode. In aggregation by maximum mode,
the price of reducing computational time leads to
over-provisioning, which causes additional energy consumption.
In the aggregation by mean mode, the price paid is the
degradation of the quality of service (QoS).  However, these
two modes still result in significant cost savings compared to
the scenario when all servers are on during the entire
duration.}  In addition, we propose two aggregation methods,
static (periodic) aggregation and dynamic (aperiodic)
aggregation. We found that in the aggregation by maximum mode,
the dynamic aggregation resulted in cost savings of up to
approximately 18\% over the static aggregation. In the
aggregation by mean mode, the dynamic aggregation by mean could
save up to approximately 50\% workload rearrangement compared
to the static aggregation by mean mode. Overall, our
methodology helps to understand the trade-off in energy-aware
aggregation.
\\ 
\end{abstract}

\section{Introduction}
Cloud computing has many advantages such as flexibility,
manageability, and scalability  \cite{Armbrust09abovethe}. Data
centers serving cloud computing face a number of challenges. A
key aspect is to minimize the total cost of ownership (TCO),
while meeting customers' workload needs. Thus, from the
perspective of  data center operators' TCO, which comprises
both energy consumption and infrastructure costs, it is
important to minimize both together. It has been reported that
the power consumption in cloud data centers has increased 400\%
over the past decade
\cite{Filani_DynamicDataCenterPowerManagement}. Data centers
consumed 61 billion kilowatt- hours of power in 2006, according
to a report of the U.S. Environmental Protection Agency (EPA)
in 2007~\cite{EPA}. That is, 1.5 percent of all power consumed
in the United States--at a cost of 4.5~billion dollars. Data
center energy costs are approaching overall hardware costs
\cite{Barroso_ThePriceOfPerformance} and even worse, continue
to increase at a rate that is faster than any others
\cite{TowardsEnegy-aware}.  The energy consumption is comprised
of multiple elements, such as servers, cooling, and power
distribution loss. We focus on the costs of servers in this
paper, i.e., the energy consumption to run servers and the
amortized capital expenditures (CAPEX) of servers.

Data center workload traces reveal that the workload is highly
dynamic
\cite{Barroso_thecase,Zhao_EnergyAwareServerProvisioning,Pinheiro02dynamiccluster}.
Due to the variety of Internet services forming workloads for
different data center operators, the workloads can be
significantly  different from one data center operator to
another. {Ideally, we want to
predict the workload by stochastic distributions or by certain
deterministic patterns by history information. However the data
center work load is highly dynamic. Namely, it is difficult to
predict for long term. To predict the workload accurately, the
workload should be predicted every once a while.} Data center
operators must provision sufficient resources to satisfy the
workload as per the service level agreements (SLA) with their
customers. If such workloads have temporal peaks, it can result
in over-provisioning for off-peak workloads. It has been found
that there is significant power consumption when the CPU is
idle, that is, at \textit{base power}
\cite{Pinheiro02dynamiccluster}. It is measured that idle
servers consume more than 66\% of the peak power
\cite{Zhao_EnergyAwareServerProvisioning,Barosso_PowerProvisioning}.
On the the other hand, for certain type of workloads, the
deadline to complete the workload may be somewhat relaxed,
which allows for evenly distributing the workload over a
shorter window of time without any additional penalty. This
situation leads to consideration for the aggregation by mean
mode rather than the aggregation by maximum mode.

\comment{Virtualization, such as, Linux VServer, Parallels,
VMware, Xen, is the technology that enables different services
to run in a virtually isolated environment and allows resources
that are allocated to these services scale up and down
transparently and seamlessly
\cite{Virtualization:usingvirtualization}. Consolidation is the
technology utilizes the capability of migration to consolidate
workload into minimum number of running servers and maximize
the number of idle servers~\footnote{``Minimum number of
running servers" does not refer to fully utilize the server
since server may not be fully utilized on purpose to guarantee
QoS.}~\cite{Zhao_EnergyAwareConsolidation}.  VMware
\textit{Distributed Power Management} (DPM) in VMware Vserver
adopts the consolidation idea to improve the power efficiency
\cite{DPM}.  The way of assigning servers is called workload
dispatching \cite{Zhao_EnergyAwareServerProvisioning}.
Assigning resource (servers) in a
Last-Released-First-Used(LRFU) manner  can minimize the needs
of consolidation \cite{Lu_SimpleAndEffective}.  That is, we use
the server that is released most recently when the workload
increases. We assume LRFU job dispatching is adopted in this
paper.}

The above suggests that an important strategy for data center
operators is to consolidate jobs (which comprise workloads)
into a minimum number of servers and switch idle servers off.
However, switching servers on and off impacts the wear-and-tear
cost and consolidation cost. It has been observed that the hard
disk is the most vulnerable part in a data center
infrastructure; the majority (78\%) of hardware
failure/replacement is due to hard disks
\cite{Vishwannath_CharacterizingCloudReliability}. To represent
the wear-and-tear costs due to switching on and off, we
amortize the CAPEX of servers by dividing the average price of
servers by the average number of switching on/off cycles of the
hard disk. On the other hand, the source server needs to run
for additional amount of time to keep the states of running
application when doing consolidation. This then leads to
consuming additional energy. Therefore, we consider two cost
components in this paper: the energy consumption cost and the
switching on/off cost. The presence of relationships between
adjacent time slots (when workloads are estimated and reviewed)
requires a global optimization framework over a temporal window
spanning several hours, which may form a planning window or a
forecast window. {In addition, the
optimal solution should be solved quickly enough, especially
for large scale data centers with short-term predictability.
Otherwise, when the solution is computed, the time window has
past.} We found that if all time slots are considered over a
planning window, the computational time to obtain the solution
is high. Thus, we propose to aggregate time slots of workload
demand to reduce computational complexity, using two different
strategics: static (periodic) and dynamic (aperiodic).

Another aspect to consider about a data center is machine
variations. It is rare that all machines at a data center are
of the same type (``homogeneous"). Often, a data center is
heterogeneous with machines of different types. More
realistically, a data center has heterogeneity but with
clusters of homogeneous machines, where machines in a cluster
get replaced about the same time from another cluster. Thus, we
wish to understand how the problem at hand is impacted when
data center machine configurations are of three possible types:
homogenous, all heterogeneous, or a homogeneous-heterogeneous
mix. While a data center with all homogeneous  or all
heterogeneous machines is unlikely in reality, we use it  for
the purpose of benchmarking and to show how we develop the
model for the more realistic mixed homogeneous-heterogeneous
data center.

\comment{We use integer programming to determine the optimal
number of running servers that minimized the cost of energy
consumption and CAPEX of servers in the entire planning
horizon.}

To summarize, our major contributions are:

\begin{enumerate}

\item To present integer linear programming (ILP) formulations to determine the optimal number of running servers over a temporal window where load adjacencies are taken into account. In particular, these formulations are presented for  three different data center configurations: homogeneous, heterogeneous, and mixed homogeneous-heterogeneous.

\item To propose two workload aggregation \textsl{methods},
    static (periodic) and dynamic (aperiodic), to reduce
    the computational time to determine the optimum. In
    addition, two workload aggregation \textsl{modes}
    (aggregation by maximum and aggregation by mean) are
    introduced to address  differing workload deadlines and
    service level agreements. To combat the pitfalls of
    static aggregations for both modes, we further enhance
    how dynamic aggregations are considered.

\item To consider workloads for a number of different
    distributions and conduct a comprehensive study to
    understand the impact on the optimal solution as well
    as on the aggregation schemes, in order to present the
    trade-off between energy-aware aggregation and the
    impact on the overall cost.

\item To present a sensitivity analysis on the optimal
    solution by varying weights of the power cost and the
    switching on-off cost components.

\end{enumerate}

The remainder of this paper is organized as follows. We present optimization formulations in Section~\ref{formulation}. Two aggregation modes, aggregation by maximum and  aggregation by mean, and two aggregation methods,  static aggregation and  dynamic aggregation, are presented in Section~\ref{aggregationmethods}. Our numeric studies are discussed in
Section~\ref{Results}. Section~\ref{related_work} summarizes the related work.
Finally, Section~\ref{Conclusion} concludes the paper.

\section{Related Work}\label{related_work}

A significant amount of work focuses on saving energy on
servers in data centers
\cite{Bertini_poweroptimization,Bianchini04powerand,Bichler_capacityplanning,Bohrer_theCaseForPowerManagement,Chen-Sigmetrics-ManagingServerEnergy,
Elnozahy02energy-efficientserver,Kusic_powerandperformance,Petrucci_dynamicoptimization_part2,Pinheiro02dynamiccluster}.
Most of the work evaluated their approaches based on a
relatively small  number of servers (around 10). While some
have considered a size of up to 100, the impact on the
computational time when a large number of servers are
considered  has not been addressed in these works.

Switching the server on and off was originally proposed by
Pinheiro \textsl{et. al.} \cite{Pinheiro02dynamiccluster} to
save energy. \cite{Pinheiro02dynamiccluster} used the
Proportional-Integral-Differential (PID) method based on
control theory to predict the demand for the next decision
point. This method takes the current demand status, the
previous accumulated demand status, and the demand change speed
into consideration and gives different weights to decide the
demand for the next decision point. Bichler \textsl{et. al.}
used a mixed integer programming (MIP) model to formulate the
capacity planning problems for virtualized servers
\cite{Bichler_capacityplanning}.

Petrucci \textsl{et. al.} considered both switching on/off and
DVFS to formulate the ``virtual server cluster configuration
problem" by MIP \cite{Petrucci_dynamicoptimization_part2}. They
proposed to devise a control loop to periodically run the
optimization problem to adapt the time-varying incoming
workload of multiple applications. Their formulation yields a
local optimum due to taking the workload of only adjacent time
periods. Chen \textsl{et. al.}
\cite{Chen-Sigmetrics-ManagingServerEnergy} formulated the
objective function, considering power consumption and the
turning on cost, (without considering the turn off cost) for
different types of constraints than the ones we identified; in
addition, they  predicted the first and second moment of the
next interval arrivals and finally calculate the SLA constraint
in terms of delay based on the $G/G/m_i$ queue. They also
proposed an approach based on control theory and a hybrid
approach of queueing and control theory as alternatives. In our
earlier work, we formulated the problem of minimizing server
operational costs in heterogeneous server cases with a dynamic
demand by binary integer programming \cite{Qian_usenix2011},
where two adaptive schemes are considered, switching servers on
and off and dynamic voltage frequency scaling (DVFS), in the
scale of 100 servers.

The relationships between optimal costs and the structure of
the workload were studied in
\cite{Lu_SimpleAndEffective,Qian_ITC2011}. In
\cite{Qian_ITC2011}, it was reported that traditional
statistical characteristics are not a good indicator for this
particular optimization problem. Instead, it proposed using
\textsl{dent} to estimate the optimal, while
\cite{Lu_SimpleAndEffective} developed algorithms to compute
the optimal by identifying the \textit{critical segments} in
the workload.

In this work, we proposed models to minimize server costs in
data centers and introduce workload aggregation methods to
combat the time complexity to compute the optimal in large
scale data centers in our earlier workshop paper
\cite{Qian_E62012}. This paper is an extension of this early
work. Compared to \cite{Qian_E62012}, there are several major
improvements. First, the study conducted is comprehensive and
is scaled up to solve large-scale problems; we also considered
the consolidation cost when switching off machines in this
study.  Because certain workload analysis suggests a daily
cycle, a sinusoidal workload pattern is additionally considered
in this paper. Furthermore, the aggregation by mean method is
also proposed since some jobs can tolerate a certain amount of
delay. Lastly, we consider sensitivity analysis on the optimal
solution by varying weights of two cost components. It is also
worth noting that the numeric study in this paper covered as
many as 5,000 servers, while most existing works consider a
small number of servers (such as up to 100 servers).

Finally, multi-time period problems have been studied over
three decades in several related areas  such as transportation
research, inventory management, and telecommunication network
capacity design
\cite{Dutta-Lim,Geary01,Johnson-Montgomery,Pioro-Medhi-book,Ukkusuri_multi-period2009,Yaged73}.
While in many of these problems, the dependency arises in the
form of constraints due to the remaining capacity of one period
being used in a subsequent period, the dependency in the  data
center resource management problem discussed here is primarily
in the form of the switching on/off cost.

\section{Optimization Problem Formulation} \label{formulation}

Consider a data center with $I$ servers. Let $\mathcal{I}$
denote the set of servers, while the cardinality of this set is
denoted by $I$, i.e., $\#(\mathcal{I}) = I$. Consider a
temporal window or forecasting window of  $\Upsilon$ hours. In
practice, we expect $\Upsilon$ to be six to eight hours,  for
which the workload can be reasonably forecasted according to
historical observations. There are many techniques for
forecasting, \textsl{e.g.},
\cite{Gmach_WorkloadAnalysisAndDemandPrediction,Verma_ServerWorloadAnalysisForPowerMinimizationUsingConsolidation}.
However, load prediction or forecasting techniques are outside
the scope of this paper. The duration of $\Upsilon$ hours for
the forecasting window is divided into $T$ equal \textsl{time
slots} or \textsl{periods}\footnote{The terms time slot and
period are used interchangeably in this paper.} and the
duration of a time slot, also referred to as \textsl{slot
size}, is $\tau=\Upsilon\cdot 60/T$ minutes. We assume that the
workload on CPU needs is forecasted at the beginning of the
entire planning window.  Servers are reconfigurable at the
beginning of each time slot, which is labeled as \textsl{review
point}s. The capacity of server $i$ is denoted by $v_i$. We
want to determine  when and how many servers to reconfigure at
review points, so that the total cost of energy consumption and
amortized server CAPEX is minimized over the entire planning
window.

\subsection{Heterogeneous Server Model}

Consider first that all servers are heterogeneous; this model
is denoted by \textsl{Model-Het}. The power consumption per
time unit of running server $i \in \mathcal{I}$ is denoted by
$c_i^p$. Binary decision variables $x_{it}$ denote 1 if server
$i$ is turned \textit{on} at review point for time slot $t$.
{The switching costs of turning server $i$ on
and off are denoted by $c_i^{s+}$ and $c_i^{s-}$, respectively.
$c_i^{s+}$ is composed by wear-and-tear cost due to turning the
server on ($c_i^{w+}$), power consumption to turn the server on
($c_i^{p+}$). Thus we have
\begin{equation}
c_i^{s+}=c_i^{w+} + c_i^{p+}. \label{turn_on_cost_components}
\end{equation}}
{And $c_i^{s-}$ is composed by wear-and-tear
cost due to turning the server off ($c_i^{w-}$), power
consumption to turn the server off ($c_i^{p-}$) and power
consumption to run additional time to sustain the states of
running applications due to consolidation in the
server($c_i^{n-}$). Thus we have
\begin{equation}
c_i^{s-}=c_i^{w-} + c_i^{p-} + c_i^{n-}.\label{turn_off_cost_components}
\end{equation}}
$x_{it}=1,0$ represents whether the states of server $i$ is
``on" and ``off" at time slot $t$, respectively. Let ``1"
represent state change while ``0" stands for no change between
adjacent time slots. Then the state change from ``on" to ``off"
can be represented by $x_{it}\cdot (x_{it}-x_{i(t-1)})$ .
Similarly, the state change from off to on can be represented
by $x_{i(t-1)}\cdot(x_{i(t-1)}-x_{it})$. Our objective is to
minimize the energy cost as well as the cost of switching
servers on/off over the planning horizon. Therefore, the
objective function is given by
\begin{eqnarray}\label{heter objective function}
F&=&\sum_{t\in \mathcal{T}}\sum_{i\in \mathcal{I}} \big(c_{i}^p\cdot x_{it}+c_{i}^{s+}\cdot x_{it}\cdot (x_{it}-x_{i(t-1)}) \nonumber \\
&& +\,c_{i}^{s-}\cdot x_{i(t-1)} \cdot (x_{i(t-1)}-x_{it})\big)
\end{eqnarray}

Next we consider constraints. The main set of constraints in
this problem is that the load  must be satisfied in every time
slot $t$. {Let $d_t$ be the
workload demand at time slot $t$.} Thus we require
\begin{equation}\label{heter demand constraint}
\sum_{i\in \mathcal{I}}v_{i}\cdot x_{it}\geq d_t,\; t=1,2,\cdots,T
\end{equation}
Note that the objective function \eqref{heter objective
function} is a  quadratic function over binary variables. We
can transform this special form of a quadratic function into a
linear function by introducing additional variables and
constraints without resorting to any approximation. We
introduce two binary variables $x_i^+(t)$ and $x_i^-(t)$ to
represent switching on and off at the review point of time slot
$t$. Specifically, $x_i^+(t)=1$ represents that server $i$ is
turned on at the review point of time slot $t$. Conversely,
$x_i^-(t)=1$ means server $i$ is turned off at the review point
for time slot $t$. $x_t^+(t)=x_t^-(t)=0$ indicates that the
state of server $i$ does not change from time slot $t-1$ to
$t$. Thus, we have
\begin{equation}\label{heter switch constraint}
x_{it}\,-\,x_{i(t-1)}\,-\,x_{it}^+ \,+\,x_{it}^-=0, \; i=1,2,\cdots,I;\, t=1,2,\cdots,T.
\end{equation}
Because $x_i^+(t)$ and $x_i^-(t)$ \textsl{cannot} both be 1 at any time slot $t$, we add the  following inequalities to enforce this requirement:
\begin{equation}\label{heter switch constraint 2}
x_{it}^+ + x_{it}^-\leq 1, \; i=1,2,\cdots,I ;\, t=1,2,\cdots,T.
\end{equation}
With the aid of $x_i^+(t)$ and $x_i^-(t)$, we now transform the original quadratic objective function \eqref{heter objective function} to the following linear function:
\begin{equation}\label{heter objective function transform}
F=\sum_{t\in \mathcal{T}}\sum_{i\in \mathcal{I}} \big(c_{i}^p\cdot x_{it}+c_{i}^{s+}\cdot x_{it}^+ + c_{i}^{s-}\cdot x_{it}^-\big).
\end{equation}
{In order to separate the energy cost ($F^p$)
and wear-and-tear cost ($F^w$), we plug
\eqref{turn_on_cost_components} and
\eqref{turn_off_cost_components} into \eqref{heter objective
function transform}. We have
\begin{equation}
F^p=\sum_{t\in \mathcal{T}}\sum_{i\in \mathcal{I}} \big(c_{i}^p\cdot x_{it}+c_{i}^{p+}\cdot x_{it}^+ + (c_{i}^{p-}+c_i^{n-})\cdot x_{it}^-\big),
\end{equation}
\begin{equation}
F^w=\sum_{t\in \mathcal{T}}\sum_{i\in \mathcal{I}} ( c_i^{w+}\cdot x_{it} ^+ + c_i^{w-}\cdot x_{it}^-),
\end{equation}
\begin{equation}
F=F^p+F^w.
\end{equation}
}

We assume that all servers are set to the off status at time
slot 0 (the beginning of the planning window); thus, we have
the initial set of conditions as follows:
\begin{equation}\label{heter initialization}
x_{i0}=0, \; i=1,\cdots, I.
\end{equation}
To summarize, in \textsl{Model-Het}, the objective is to minimize \eqref{heter objective function transform}, which is subject
to \eqref{heter demand constraint}, \eqref{heter switch constraint}, and
\eqref{heter switch constraint 2}, where
 the initial conditions are given by \eqref{heter initialization} and all variables are binary variables.

\subsection{Homogeneous Server Model}\label{sec: Hom}

In the case of the homogeneous server configuration, all
servers are considered to be identical. This model is denoted
by \textsl{Model-Hom}. Although the heterogeneous model
introduced in the previous section is generic to be applied to
this model, we present a different formulation where the number
of variables and constraints are reduced significantly. Let
$c^p$ be the power consumption of running a server in a time
slot. Let $y_t$ denote the number of \textit{running}
homogeneous servers at time slot $t$. Therefore, we can reduce
the $I$ binary variables in Model-Het to a single integer
variable for each time slot in \textsl{Model-Hom}. The power
consumption for each server per time slot is denoted by $c^p$.
{The cost of switching one server on and off is
denoted by $c^{s+}$ and $c^{s-}$, respectively. $c^{s+}$ is
composed by wear-and-tear cost due to turning one server on
($c^{w+}$), power consumption to turn one server on ($c^{p+}$).
Thus we have
\begin{equation}
c^{s+}=c^{w+} + c^{p+}. \label{homo_turn_on_cost_components}
\end{equation}}
{And $c^{s-}$ is composed by wear-and-tear cost
due to turning one server off ($c^{w-}$), power consumption to
turn one server off ($c^{p-}$) and power consumption to run
additional time to sustain the states of running applications
due to consolidation in one server ($c^{n-}$). Thus we have
\begin{equation}
c^{s-}=c^{w-} + c^{p-} + c^{n-}.\label{homo_turn_off_cost_components}
\end{equation}}

\comment{Performing workload dispatching is the overhead in
this model. Because to achieve this model, we need to use the
resource in LRFU manner. That is, the latest released resource
should be used first when receiving new requests. Implementing
workload dispatching algorithms is out of the scope of this
paper. It has been studied in
\cite{Zhao_EnergyAwareServerProvisioning,
Lu_SimpleAndEffective}. However we need to keep in mind that
this overhead happens for the homogeneous server model as well
as heterogeneous homogeneous server cluster model which will be
introduced in Sec~\ref{sec: HH}.}

Similar to Model-Het, the first constraint is the workload requirement using the new variables $y_t$:
\begin{equation}
v_{t}\cdot y_{t}\geq d_t,\;t=1,\cdots,T. \label{homo demand constraint}
\end{equation}
Let $y_t^+$ denote the number of servers that is switched on at
the review point of time slot $t$. Then $y_t^+$ should take the
maximum between 0 and $y_t-y_{t-1}$. That is,
\begin{equation}
y_t^+=\max\{0, y_{t}-y_{t-1}\},\; t=1,\cdots, T. \label{homo switch on}
\end{equation}
Let $y_t^-$ be the number of servers that is switched off at the review point of time slot $t$. Similar to \eqref{homo switch on}, we have
\begin{equation}
y_t^-=\max \{0, y_{t-1}-y_t\},\; t=1,\cdots,T.\label{homo switch off}
\end{equation}
Note that \eqref{homo switch on} and \eqref{homo switch off} are not directly usable constraints.
Because we are considering a minimization problem with the cost coefficient being non-negative, we can substitute \eqref{homo switch on} by following two linear inequalities:
\begin{equation}
y_t^+\geq y_t-y_{t-1},\; t=1,\cdots, T. \label{homo switch on 1}
\end{equation}
\begin{equation}
y_t^+\geq 0,\; t=1, \cdots, T. \label{homo switch on 2}
\end{equation}
Likewise, we can substitute \eqref{homo switch off} by following two linear inequalities:
\begin{equation}
y_t^-\geq y_{t-1}-y_{t},\; t=1,\cdots, T. \label{homo switch off 1}
\end{equation}
\begin{equation}
 y_t^-\geq 0,\; t=1,\cdots, T.\label{homo switch off 2}
\end{equation}
The final set of constraints is on the number of running servers should not be larger than total number of servers
\begin{equation}
y_t\leq I,\; t=1,\cdots, T.
\end{equation}
Because it is a  minimization problem with the cost coefficient being non-negative, this constraint can  be omitted when solving the optimization problem. Since all servers are off at the beginning of the planning window (i.e., at the review point for the beginning time slot~0), we have the initial condition
\begin{equation}
y_0=0. \label{homo initialization}
\end{equation}
The objective is to minimize the total energy and switching on/off cost, which is given by
\begin{equation}\label{homo objective function}
F=\sum_{t\in \mathcal{T}}({c^p}\cdot y_t+c_t^{s+}\cdot y_t^{+}+c_t^{s-}\cdot y_t^-).
\end{equation}
{In order to separate the energy cost ($F^p$)
and wear-and-tear cost ($F^w$), we plug
\eqref{homo_turn_on_cost_components} and
\eqref{homo_turn_off_cost_components} into \eqref{homo
objective function}. We have
\begin{equation}
F^p=\sum_{t\in \mathcal{T}}\big(c^p\cdot y_t+c^{p+}\cdot y_t^+ + (c^{p-}+c^{n-})\cdot y_t^-\big),
\end{equation}
\begin{equation}
F^w=\sum_{t\in \mathcal{T}}( c^{w+}\cdot y_t ^+ + c^{w-}\cdot y_t^-),
\end{equation}
\begin{equation}
F=F^p+F^w.
\end{equation}
}
To summarize, in the Model-Hom, we  minimize \eqref{homo
objective function}, which is subject to \eqref{homo demand
constraint}, \eqref{homo switch on 1}, \eqref{homo switch on
2}, \eqref{homo switch off 1}, and \eqref{homo switch off 2}
and the initialization condition is given by \eqref{homo
initialization}. This is an integer linear programming (ILP)
problem. \comment{ For sake of clarity, we summarize this model
in \ref{summary of Model-Hom}
\begin{figure}[!t]\label{summary of Model-Hom}
\fbox{
\parbox[t]{.45\textwidth}{
\textbf{constants}\\
\hspace*{2em}$c^p\;$ energy consumption per server per time slot\\
\hspace*{2em}$c^{s+}\;$ cost of switching a server on\\
\hspace*{2em}$c^{s-}\;$ cost of switching a server off\\
\hspace*{2em}$v\;$ capacity of a server\\
\hspace*{2em}$d_t\;$ demand at $t$\\
\hspace*{2em}$y_0=0\;$ initial (time slot $0$) state of servers \\
\textbf{variables}\\
\hspace*{2em}$y_{t}\in\mathcal{N}\;$  \# of running servers at time slot $t$
\hspace*{2em}$y_{t}^+\in\mathcal{N}\;$  \# servers turned on at time slot $t$\\
\hspace*{2em}$y_{t}^-\in\mathcal{N}\;$ \# of servers turned off at time slot $t$\\
\textbf{objective}\\
\hspace*{2em}minimize $F=\sum_{t} (c^p\cdot y_{t}+ c^{s+}\cdot y_{t}^+ + c^{s-}\cdot y_{t}^-)$\\
\textbf{constraints}\\
\hspace*{2em}$v_{t}\cdot y_{t}\geq d_t,\;\forall t\in\mathcal{T}$\\
\hspace*{2em}$y_{t}\leq I, \;\forall t\in\mathcal{T}$\\
\hspace*{2em}$y_t^+\geq 0, \;\forall t\in\mathcal{T}$\\
\hspace*{2em}$y_t^+\geq y_t-y_{t-1}, \;\forall t\in\mathcal{T}$\\
\hspace*{2em}$y_t^-\geq 0, \;\forall t\in\mathcal{T}$\\
\hspace*{2em}$y_t^-\geq y_{t-1}-y_{t}, \;\forall t\in\mathcal{T}$\\
}
}
\caption{\textsl{Model-Hom}: IP - Homogeneous Server Case}
\end{figure}
}

\subsection{Heterogeneous Homogeneous-Server-Cluster Model}\label{sec: HH}

In a  data center, it is more realistic that there are
different clusters of servers and each cluster has a certain
number of homogeneous servers while servers may be different
from one cluster to another, rather then all being homogeneous
or all being heterogeneous. Thus, this is a hybrid of the afore
mentioned configuraitons. This model can also be used when
homogenous servers are required to be partitioned into multiple
clusters for ease of management. We denote this model by
\textsl{Model-HH}. Denote the set of clusters by $\mathcal{J}$
and its cardinality, $J = \#(\mathcal{J})$, where $1\leq J\leq
I$, represents the number of clusters. The two  models
presented so far correspond to the two extremes in this model:
when $J=1$, it is Model-Hom; when $J=I$, it becomes Model-Het.

Let the energy consumption of running a server in cluster $j$
at a time slot be $z_j^p$. Denote the number of servers in
cluster $j$ by $I_j$, where $\sum_j I_j = I$. We denote the set
of the number of running servers for cluster $j$ by
$\mathcal{N}_j$, which can be $0,1,\cdots, I_j$. Let $z_{jt}$
be the number of running servers in cluster $j$ at time slot
$t$. \comment{As we have mentioned in Sec.~\ref{sec: Hom}, the
workload need to be dispatched in the LRFU manner.} The costs
of switching a server in cluster $j$ on and off are represented
by $c_j^{s+}$ and $c_j^{s-}$, respectively.
{$c_j^{s+}$ is composed by wear-and-tear cost
due to turning the server in cluster $j$ on ($c_j^{w+}$), power
consumption to turn the server in cluster $j$ on ($c_j^{p+}$).
Thus we have
\begin{equation}
c_j^{s+}=c_j^{w+} + c_j^{p+}. \label{HH_turn_on_cost_components}
\end{equation}
And $c_j^{s-}$ is composed by wear-and-tear cost due to turning
the server in cluster $j$ off ($c_j^{w-}$), power consumption
to turn the server in cluster $j$ off ($c_j^{p-}$) and power
consumption to run additional time to sustain the states of
running applications due to consolidation in the server of
cluster $j$ ($c_j^{n-}$). Thus we have
\begin{equation}
c_j^{s-}=c_j^{w-} + c_j^{p-} + c_j^{n-}.\label{HH_turn_off_cost_components}
\end{equation}
}

We next introduce constraints in this problem. First, the
workload requirements need to be satisfied all the time:
\begin{equation}
\sum_{j\in\mathcal{J}}v_{jt}\cdot z_{jt}\geq d_t,\; t=1,\cdots, T. \label{HH demand constraint}
\end{equation}
Secondly, the number of running servers cannot be larger than
the total number of servers in that cluster:
\begin{equation}
z_{jt}\leq I_j, \;j=1,\cdots, J;\,t=1,\cdots, T. \label{HH cluster capacity}
\end{equation}
The third set of constraints is similar to \eqref{homo switch
on} and \eqref{homo switch off} in Model-Hom. But they need to
be applied to each of the $J$ clusters. Let $z_{jt}^+$ be the
number of servers turned on in cluster $j$ at the review point
of time slot $t$ while $z_{jt}^-$ is the number of servers
turned off in cluster $j$ at the review point of time slot $t$.
We use the same technique as in \textsl{Model-Hom} to transform
the constraints to linear constraints
\begin{equation}\label{HH switch on 1}
z_{jt}^+\geq 0, \; j=1,\cdots, J;\,t=1,\cdots, T.
\end{equation}
\begin{equation}\label{HH switch on 2}
z_{jt}^+\geq z_{jt}-z_{j(t-1)},\; j=1,\cdots, J;\,t=1,\cdots, T.
\end{equation}
\begin{equation}\label{HH switch off 1}
z_{jt}^-\geq 0, \; j=1,\cdots, J;\,t=1,\cdots, T.
\end{equation}
\begin{equation}\label{HH switch off 2}
z_{jt}^-\geq z_{j(t-1)}-z_{jt}, \; j=1,\cdots, J;\,t=1,\cdots, T.
\end{equation}
And the objective function is given by
\begin{equation}\label{HH objective function}
F=\sum_{t\in\mathcal{T}}\sum_{j\in\mathcal{J}} (c_j^p\cdot z_{jt}+ c_j^{s+}\cdot z_{jt}^+ + c_j^{s-}\cdot z_{jt}^-)
\end{equation}
{Similar to previous two models, we separate
the energy cost ($F^p$) and wear-and-tear cost($F^w$) and we
have
\begin{equation}
F^p=\sum_{t\in \mathcal{T}}\sum_{j\in \mathcal{J}} \big(c_{j}^p\cdot z_{jt}+c_j^{p+}\cdot z_{jt}^+ + (c_j^{p-}+c_j^{n-})\cdot z_{jt}^-\big),
\end{equation}
\begin{equation}
F^w=\sum_{t\in \mathcal{T}}\sum_{j\in \mathcal{J}} ( c_j^{w+}\cdot z_{jt} ^+ + c_j^{w-}\cdot z_{jt}^-),
\end{equation}
\begin{equation}
F=F^p+F^w.
\end{equation}
}
Because we assume that at the beginning of the planning
period, all servers are off, we have
\begin{equation}\label{HH initialization}
z_{j0}=0, \; j=1,\cdots, J.
\end{equation}
Here, the objective is to minimize \eqref{HH objective function} which is subject to \eqref{HH demand constraint}, \eqref{HH cluster capacity}, \eqref{HH switch on 1}, \eqref{HH switch on 2}, \eqref{HH switch off 1}, and \eqref{HH switch off 2}, and the initial conditions are given by \eqref{HH initialization}.

\section{Aggregating Demand to Reduce Computational Time} \label{aggregationmethods}

The number of variables for the heterogeneous case, the
homogenous case, the heterogeneous homogeneous-server-cluster
case, presented in Section~\ref{formulation} are $3\times
I\times T$, $3\times T$, and $3\times J \times T$,
respectively. Since the computational time grows with the
number of  variables for these models (to be discussed further
in Section~\ref{sec:sub:CompTime}), the time complexity for the
heterogeneous and heterogeneous homogeneous-server-cluster
cases differs significantly for large scale data centers (large
$I$ and $J$, respectively) compared to the homogenous case.

To reduce the computational time, we need to either reduce the
number of variables in the original problem or develop a
heuristic to approximate the original problem. In this paper,
we propose an aggregation heuristic strategy in which a certain
number of contiguous time slots are combined into a single
aggregated slot to reduce the total number of time slots for
the workload, by considering load affinity. The value of the
workload demand on an aggregated time slot is decided by the
workload of the original time slots and the service-level
agreement (SLA).

\subsection{Aggregation by Maximum}
If an SLA stringently requires that the workload should be
satisfied all the time, then the workload of an aggregated time
slot takes the maximum of the demand of the original time
slots. For example, we want to aggregate contiguous slots
$[d_{k}, d_{k+1},\cdots, d_{k+\ell-1}]$ into an aggregated time
slot; then the new demand $\hat{d}_k$ over the $\ell$ times of
the original slot size is given by
\begin{equation}
{d}_{k}^{\max} = \max \{d_{k}, d_{k+1},\cdots,d_{k+\ell-1}\}.
\end{equation}
This method introduces an artificial increase in the demand, which in turn, causes extra consumption of power energy. On the other hand, aggregation smoothes out the irregularity of workload, which  affects  the switching cost. This raises the issue of trading off the computing time of  running the model at the expense of extra cost on energy consumption.

\subsubsection{\underline{Static Aggregation by Maximum}}

In the static approach (\figureref{max_illustration}(a)), we
aggregate every $M$ contiguous  time slots into one, i.e., the
aggregation window is periodic. We have $\hat{T}=\lceil T/M
\rceil$, where $\hat{T}$ denotes the reduced number of time
slots. The slot size of the aggregated workload (except the
last slot) is $M$ times  the slot size of the original demand.
The slot size of the last slot is $T-M\times (\hat{T}-1)$. The
aggregated workload ${\lceil t/M\rceil}$ is given by
\begin{equation}
{d}_{\lceil t/M\rceil}^{\max}= \max\{d_{(\lceil t/M\rceil-1)\cdot M +1},\cdots,d_{(\lceil t/M \rceil-1)\cdot M+M}\} \label{static_aggregated demand}
\end{equation}
\begin{figure}[!t]
\centering
\epsfxsize=.5\textwidth
\epsffile[97   239   514   552]{./figures/max_static_dynamic_illustration.eps}
\caption{Aggregating Workload: Static (periodic) vs. Dynamic (aperiodic)}  \label{max_illustration} \end{figure}

\subsubsection{\underline{Dynamic Aggregation by Maximum}}

Our dynamic aggregation approach
(\figureref{max_illustration}(b)) improves on the static
(periodic) approach, with a goal to improve the overall cost.
\comment{Although some waste of energy consumption is
inevitable, we can alleviate waste by using dynamic
aggregation.}  Instead of aggregating the workload in  fixed
numbers of original time slots statically, the adaptive
aggregation method aggregates an arbitrary number of time slots
(aperiodic) as long as the number of aggregated time slots is
$\hat{T}$ such that the sum of the difference between the
aggregated workload and the original workload is minimized.
That is, we seek to minimize
\begin{equation}\label{max_dynamic_aggregation_objective}
\sum_{k=1}^{\hat{T}} \sum_{\ell=1}^{T_k}(d_k^{\max}-d_{k\ell}), \mbox{ where } d_k^{\textrm{max}}= \max \{d_{k1}, \cdots, d_{kT_k}\},
\end{equation}
which is subject to
\begin{equation}
\sum_{k=1}^{\hat{T}} T_k = T. \label{aggregation_target}
\end{equation}

To this end, we need to choose $\hat{T}-1$ review points out of
$T-1$, requiring ${{T-1}\choose{\hat{T}-1}}$ operations. This
extra computational time complexity is contradictory to the
purpose of doing aggregation. Thus, we propose \textsl{local
smooth} heuristics to implement this idea. We aggregate the
time slots that are locally ``smoother" together. In order to
do this, we first define the smooth index of workloads. The
smooth index is the absolute value of the difference between
the workload demand of adjacent time slots (adjacent workloads
for short). Therefore, we obtain $T-1$ smooth indices. Then we
pick the smallest non-zero smooth index and compare two
adjacent workloads associated with this smooth index. The
smaller workloads are aggregated into a maximum workload over
these slots; the slot size of the aggregated workload is the
sum of two slot sizes of adjacent slots and the new workload
for the aggregated slot takes the maximum of these demands. The
smooth index is updated for the new aggregated demand series.
We repeat this procedure until the target number of slots is
reached. We call this procedure  the \textit{local smooth
algorithm} (see Algorithm~\ref{local_smooth_algorithm}). Denote
``InMin" to be the procedure to find the index of minimum value
in a vector, ``Max" be the procedure to find the maximum value
while "Mean" be the procedure taking the average, which is used
in the aggregation by mean mode. The value of the workload in
time slot $i$ is denoted by $d[i]$.

It is possible to make further improvements to
Algorithm~\ref{local_smooth_algorithm}. For this, denote the
smooth index vector by $si$. Scalar $ap$ records the
aggregation point. Vector $ss$, which is initialized to $T$
ones, records the size of each time slot.  The \textit{improved
local smooth algorithm} is presented in
Algorithm~\ref{local_smooth_algorithm_improved}.
\begin{algorithm}
\caption{Local Smooth Algorithm}
\label{local_smooth_algorithm}
\begin{algorithmic}[1]
\STATE $j \gets T$ \hfill {Initialization}
\WHILE {$j > \hat{T}$}
        \FOR{$i:= 2 \to j$}
          \STATE $si[i-1] \gets |d[i]-d[i-1]|$ \hfill Compute smooth index
        \ENDFOR
        \STATE $ap \gets$ InMin(si) \hfill Find aggregation point
        \STATE $d[ap] \gets \textrm{Max/Mean}(d[ap], d[ap+1])$
        \STATE $d[ap+1] \gets \textrm{Max/Mean}(d[ap], d[ap+1])$ \hfill Aggregate
        \STATE $ss[ap] \gets ss[ap] +ss[ap+1] $ \hfill Compute the size of aggregated slots
       \IF{$ap\neq j-1$} 
           \FOR {$k: = ap+1 \to j-1$}
           \STATE $d[k] = d[k+1]$ \hfill Adjust the index of slots behind the aggregation point
           \STATE $ss[k] = ss[k+1]$ \hfill Adjust the corresponding slot size
           \ENDFOR
        \ENDIF
        \STATE $j\gets j-1$ \hfill Decrease the number of slots by 1
\ENDWHILE
\RETURN d, ss
\end{algorithmic}
\end{algorithm}

\begin{algorithm}
\caption{Improved Local Smooth Algorithm}
\label{local_smooth_algorithm_improved}
\begin{algorithmic}[1]
\STATE $j \gets T$ \hfill {Initialization}
\FOR{$i:= 2 \to j$}
          \STATE $si[i-1] \gets (d[i]-d[i-1])$ \hfill Compute smooth index
\ENDFOR
\WHILE {$j > \hat{T}$}
        \STATE $ap \gets$ InMin($|$si$|$) \hfill Find aggregation point
        \STATE $d[ap] \gets \textrm{Max/Mean}(d[ap], d[ap+1])$
        \STATE $d[ap+1] \gets \textrm{Max/Mean}(d[ap], d[ap+1])$ \hfill Aggregate
        \STATE $si[ap-1] \gets d[ap]-d[ap-1]$ \hfill Update smooth index
        \STATE $si[ap] \gets d[ap+2]-d[ap+1]$  \hfill Update smooth index
        \STATE $ss[ap] \gets ss[ap] +ss[ap+1] $ \hfill Compute the size of aggregated slots
       \IF{$ap\neq j-1$} 
           \FOR {$k: = ap+1 \to j-1$}
           \STATE $d[k] \gets d[k+1]$ \hfill Adjust the index of slots behind the aggregation point
           \STATE $ss[k] \gets ss[k+1]$ \hfill Adjust the corresponding slot size
              \IF {$k< j -1$}
                  \STATE $si[k] \gets si[k+1] $ \hfill Adjust the smooth index
              \ENDIF
           \ENDFOR
        \ENDIF
        \STATE $j\gets j-1$ \hfill Decrease the number of slots by 1
\ENDWHILE
\RETURN d, ss
\end{algorithmic}
\end{algorithm}

It is easy to see that Algorithm~\ref{local_smooth_algorithm}
has complexity $\mathcal{O}{(T^2)}$, while
Algorithm~\ref{local_smooth_algorithm_improved} has linear
complexity  $\mathcal{O}{(T)}$, which is achieved by
recomputing two smooth indexes adjacent to the selected
aggregation point only. \comment{However, since it is based on
the local information, there is no guarantee that this method
will reach the global optimal solution. In some rare cases, it
is possible that this solution is worse than static allocation
in terms of the optimum.} Note that the computational time for
the dynamic aggregation has no significant difference with its
static aggregation counterpart since they have the same number
of variables and constraints. To differentiate between the
proposed dynamic aggregation and the implemented dynamic
aggregation, we call the proposed dynamic aggregation as the
\textsl{strict} dynamic aggregation.

\subsection{Aggregation by Mean}
The workload demand of the aggregated time slot  can also take a certain percentile of the workload demand of the original slots. Note that \textsl{Aggregation by maximum}  uses the workload to be the 100th percentile of the workload of the original slots. For many long-lived jobs that do not need to be executed in real-time, such as data warehousing or scientific computing, the workload can be arranged over time as long as the average workload over an acceptable time window is completed. Thus, we also introduce the mode of  aggregation by mean: the workload demand of the aggregated time slot takes the mean of the workload demand of the original slots. Consider aggregating loads for contiguous slots $[d_{k}, d_{k+1},\cdots, d_{k+\ell-1}]$ into one time slot; then, the new demand $\bar{d}_k$ with $\ell$ times of the original slot size is given by
\begin{equation}
\bar{d}_{k} = \textrm{mean} \{d_{k}, d_{k+1},\cdots,d_{k+\ell-1}\}.
\end{equation}
Compared to the aggregation by maximum mode, the aggregation by mean mode does not introduce the artificial increase of the workload demand and smoothes out the irregularity of the workload.

\subsubsection{\underline{Static Aggregation by Mean}}

As before, we aggregate every $M$ contiguous time slots into 1. The only difference is that the aggregated workload takes the average of the original workload:
\begin{equation}
\bar{d}_{\lceil t/M\rceil}= \textrm{mean}\{d_{(\lceil t/M\rceil-1)\cdot M +1},\cdots,d_{(\lceil t/M \rceil-1)\cdot M+M}\} \label{static_mean_aggregated demand}
\end{equation}
The application of this method is based on the ability to re-arrange user requests within a certain time window, which is a subset of the planning window, for some applications that do not require real-time execution. In  other words, the requested load can be either executed in advance or delayed in the data center. An example of static aggregation by average is shown in \figureref{average_illustration}.

\begin{figure}[!t]
\centering
\epsfxsize=.5\textwidth
\epsffile[97   229   514   562]{./figures/mean_static_dynamic_illustration.eps}
\caption{Illustration of Aggregating Workloads by Mean}  \label{average_illustration} \end{figure}

\subsubsection{\underline{Dynamic Aggregation by Mean}}

Similar to the previous instance, but this time to avoid delay or advance workload as much as possible, we also propose the counterpart of the dynamic aggregation  by max. The objective function is to minimize
\begin{equation}
\sum_{k=1}^{\hat{T}}\sum_{\ell=1}^{T_k}|d_{k\ell}-\bar{d}_k|, \mbox{ where } \bar{d}_k = \textrm{mean}\{d_{k1},\cdots,d_{kT_k}, \}\label{mean_dynamic_aggregation_objective}
\end{equation}
which is subject to \eqref{aggregation_target}.

\figureref{average_illustration}(b) illustrates the dynamic
aggregation by average. It is worth noting that although
\eqref{max_dynamic_aggregation_objective} and
\eqref{mean_dynamic_aggregation_objective} look similar, the
objectives of dynamic aggregation  by maximum and dynamic
aggregation by mean are different.  Aggregation by maximum aims
to reduce cost due to energy waste while aggregation by mean
targets to reduce the movement of the workload. Thus, compared
to the static aggregation,  the aggregation by maximum results
in less energy cost while in aggregation by mean, the energy
cost is always the same. However, the proposed dynamic
aggregation method has no constraints on the number of original
slots  combined to create an aggregated slot. In practical
applications, the workload can only be executed in advance or
delayed up to a certain time. Let $S$ be the maximum number of
continuous time slots that can be aggregated to meet the
delayed requirement. Consequently, this problem is also subject
to
\begin{equation}
\max\{s_1,\cdots,s_{\hat{T}}\}\leq S.
\end{equation}
The exact solution based on improved local smooth algorithms
requires $n!$ time. We propose an approximation scheme with low
complexity that relaxes the target number of aggregated
workload slots to guarantee that the movement of the workload
is less than a certain threshold. The modified algorithm is
shown in
Algorithm~\ref{local_smooth_algorithm_improved_with_constraints}.
To illustrate, \figureref{average_illustration}(c) presents an
example of dynamic aggregation constrained by $S=8$. A problem
with this implementation is that it may not have a feasible
solution. To address this issue, we swap lines~22 and 23 in
Algorithm~\ref{local_smooth_algorithm_improved_with_constraints}
to relax the target number of time slots to guarantee that
there is a solution.
\begin{algorithm}
\caption{Improved Local Smooth Algorithm with Constraints on Advance and Delay }
\label{local_smooth_algorithm_improved_with_constraints}
\begin{algorithmic}[1]
\STATE $j \gets T$ \hfill {Initialization}
\FOR{$i:= 2 \to j$}
          \STATE $si[i-1] \gets (d[i]-d[i-1])$ \hfill Compute smooth index
\ENDFOR
\WHILE {$j > \hat{T}$}
\IF{$ss[ap] + ss[ap+1] \leq S$}
        \STATE $ap \gets$ InMin($|$si$|$) \hfill Find aggregation point
        \STATE $d[ap] \gets \textrm{Mean}(d[ap], d[ap+1])$
        \STATE $d[ap+1] \gets \textrm{Mean}(d[ap], d[ap+1])$ \hfill Aggregate
        \STATE $si[ap-1] \gets d[ap]-d[ap-1]$ \hfill Update smooth index
        \STATE $si[ap] \gets d[ap+2]-d[ap+1]$  \hfill Update smooth index
        \STATE $ss[ap] \gets ss[ap] +ss[ap+1] $ \hfill Compute the size of aggregated slots
       \IF{$ap\neq j-1$} 
           \FOR {$k: = ap+1 \to j-1$}
           \STATE $d[k] \gets d[k+1]$ \hfill Adjust the index of slots behind the aggregation point
           \STATE $ss[k] \gets ss[k+1]$ \hfill Adjust the corresponding slot size
              \IF {$k< j -1$}
                  \STATE $si[k] \gets si[k+1] $ \hfill Adjust the smooth index
              \ENDIF
           \ENDFOR
        \ENDIF
        \STATE $j\gets j-1$ \hfill Decrease the number of slots by 1
        \ENDIF
\ENDWHILE
\RETURN d, ss
\end{algorithmic}
\end{algorithm}

\section{Results and Discussions} \label{Results}

Before we discuss our results, we summarize a few key points of
the approaches presented so far. The aggregation methods are
proposed to reduce the computation time of the workload
planning problem for large scale data centers. In practice, if
the workload cannot be rearranged over time, aggregation by
maximum should be used; otherwise, aggregation by mean can be
adopted. The price of aggregation by maximum is
over-provisioning which causes extra energy consumption. The
price of aggregation by mean is the workload rearrangement.
Dynamic aggregation is proposed to alleviate over-provisioning
and workload rearrangement in aggregation by maximum and by
mean, respectively. In this section, we quantitatively study
the pros and cons of proposed aggregation methods.

\subsection{Experiment Setup}
In our study, the server's CPU frequency set and power
consumptions are adopted from
\cite{Chen-Sigmetrics-ManagingServerEnergy}, except that we use
the maximum frequency only. For ease of comparison, the
capacity of each server is normalized to 1. Greenberg
\textsl{et. al.} \cite{Greenberg_thecostofcloud} use \$.07 per
killowatt-hours (kWh) as the utility price. We assume the same
utility price. The server energy consumption in a time slot is
the product of the power consumption, the utility price, and
the slot size. \comment{Table~\ref{Tab:server_parameter}
presents the specifications of server CPU.} We use server CPU
frequency to be 2.6GHz, power consumption to be 100 watts with
a power cost of \$0.07 per kWh.

{Google reported
\cite{Barroso_theDatacenterAsaComputer} that the hard disk is
the most vulnerable part in a server and the personnel cost for
each repair is \$100 and the replacement cost is 10\% of the
server cost (\$2,000)}. We assume the lifetime for a disk to be
60,000 switching-on-and-off cycles. Using this, we arrive at
0.5 cents for the wear-and-tear cost per switching-on-off
cycle. Out of 0.5 cents, wear-and-tear cost due to switching on
is usually higher than that due to switching off; thus, we
split 0.5 cents to 0.3 cents and 0.2 cents for wear-and-tear
cost due to switching on ($c_i^{w+}/c^{w+}/c_j^{w+}$) and
wear-and-tear cost due to switching off (
$c_i^{w-}/c^{w-}/c_j^{w-}$). Since turning on draws much more
power than turning off in most cases we assume that the power
consumption for switching on ( $c_i^{p+}/c^{p+}/c_j^{p+}$) and
switching off ( $c_i^{p-}/c^{p-}/c_j^{p-}$) is 0.02 cents and
0.005 cents, respectively.  This analysis of the turning on/off
cost is similar to that in
\cite{Chen-Sigmetrics-ManagingServerEnergy} except that we
differentiate the cost of switching on and off. The cost of
switching on ($c_i^{s+}/c^{s+}/c_j^{s+}$) is 0.32 cents . When
consolidation is performed, the source server needs to run for
an additional amount of time to sustain the state of running
applications. We assume the average extra time to be 77
seconds. We arrive at 0.015 cents per server switching off
($c_i^{n-}/c^{n-}/c_j^{n-}$) as the cost of consolidation .
Because the consolidation cost is not considered in our
previous work \cite{Qian_usenix2011}, the cost of switching off
( $c_i^{s-}/c^{s-}/c_j^{s-}$) is 0.22 cents in lieu of 0.205
cents in \cite{Qian_usenix2011}.
\comment{
\begin{table}[!t]
\caption{Specification of Server CPU}\label{Tab:server_parameter}
\centering
\vspace{-.2in}
\footnotesize{
\begin{tabular}[t]{c|c|c|c}
\hline
Frequency&Power Consumption&Power Cost&Normalized Capacity\\
\hline
2.6GHz&100watts&.7$\tau$&1\\
\hline
\end{tabular}
}
\end{table}
}

We next consider the scenario that the cost components may
change due to the fluctuation of utility price and technology
advancements. We define a cost model to consider this factor
for cost sensitivity analysis. We weight the utility price by
$\beta$ and the wear-and-tear cost by $1-\beta$.  We define the
cost of running a 100-watt server for 5 minutes as $\beta\cdot
0.14 \cdot (100/1000) \cdot 5/60$. The wear-and-tear cost due
to switching on is defined by $(1-\beta)\cdot 0.6$. The
wear-and-tear cost due to switching off is defined by
$(1-\beta)\cdot 0.4$. Therefore, the power consumption of
switching a server on and off is given by $\beta \cdot 0.04$
and $\beta \cdot 0.01$, respectively. The power consumption to
do consolidation when switching off a server is given by $\beta
\cdot 0.03$. Thus, we have the cost of $\beta\cdot 7/60$ for
running a 100 watt server for 5 minutes, the cost of
$0.6-0.56\beta$ for switching a server on and $0.4-0.36\beta$
for switching a server off. {$\beta=0.5$ is used for the cost model in all other cases in
our study.}

We assume that the utilization of the data center is 20\%.
Therefore, the average workload to the cloud is $I\times 0.2$.
We also assume that the workload can be  forecasted and
profiled every 5 minutes. Due to the diurnal behavior
associated with human beings' working cycles, we chose the 8
hour work time as the planning horizon where the dynamically
changing workload from one time slot to another is generated
for our study. The sinusoidal function and three different
random distributions with the same average are used to generate
temporally dynamic workload profiles. The three random
distributions are Erlang-2 (smooth), exponential, and two-state
hyper-exponential (bursty).  Each random distribution is
generated 101 times using 101 independent random streams. Note
that for the workload generated by these distributions  that
are over the maximum capacity, i.e., $I$, we truncate the
maximum workload to $I$ to make the problem feasible. We also
wish to study the workload that can be represented by certain
deterministic cyclic functions. Assume that a day's workload
can be represented by a full cycle of the sinusoidal function
and the 8 hour workload window is in the range of 0 degrees and
120 degrees.  We first generate the value given by the plain
sinusoidal function in the range of 0 degree and 120 degree:
\begin{equation}
\tilde{d}_t=\sin(t\times2\times\pi/3/96), \forall t=1,\cdots,96.\label{plain sinusoidal Workload}
\end{equation}
Then, the 8 hour workload load demand with an average of $0.2\times
I$ is given by:
\begin{equation}
d_t=(\tilde{d}_t - \sum_{t\in\mathcal{T}}\tilde{d}_t/96 + 1)\times 0.2\times I, \forall t=1,\cdots,96. \label{sinusoidal Workload}
\end{equation}
Compared to three workloads generated by random distributions, since this workload is deterministic, we call it the deterministic sinusoidal workload.

\comment{\figureref{sinusoid_illustration} presents an example of the 8
hour sinusoidal workload derived from \eqref{sinusoidal Workload}
given $I=100$.
\begin{figure}[htbp]
\centering
\epsfxsize=.4\textwidth
\epsffile[96   239   514   552]{./figures/sinusoid_workload.eps}
\caption{Sinusoidal Workload when $I=100$}  \label{sinusoid_illustration}
\end{figure}
}

We ran the optimization model using CPLEX through Matlab on  an Intel(R) Core(TM)2 Duo CPU U9400 \@1.40GHz with 4GB memory.

\subsection{Computational Time of Different Models and Different Number of Workload Slots} \label{sec:sub:CompTime}

\begin{figure}[!t]
\begin{minipage}{0.48\linewidth}
\centering
\epsfxsize=\textwidth
\epsffile[95   157   515   634]{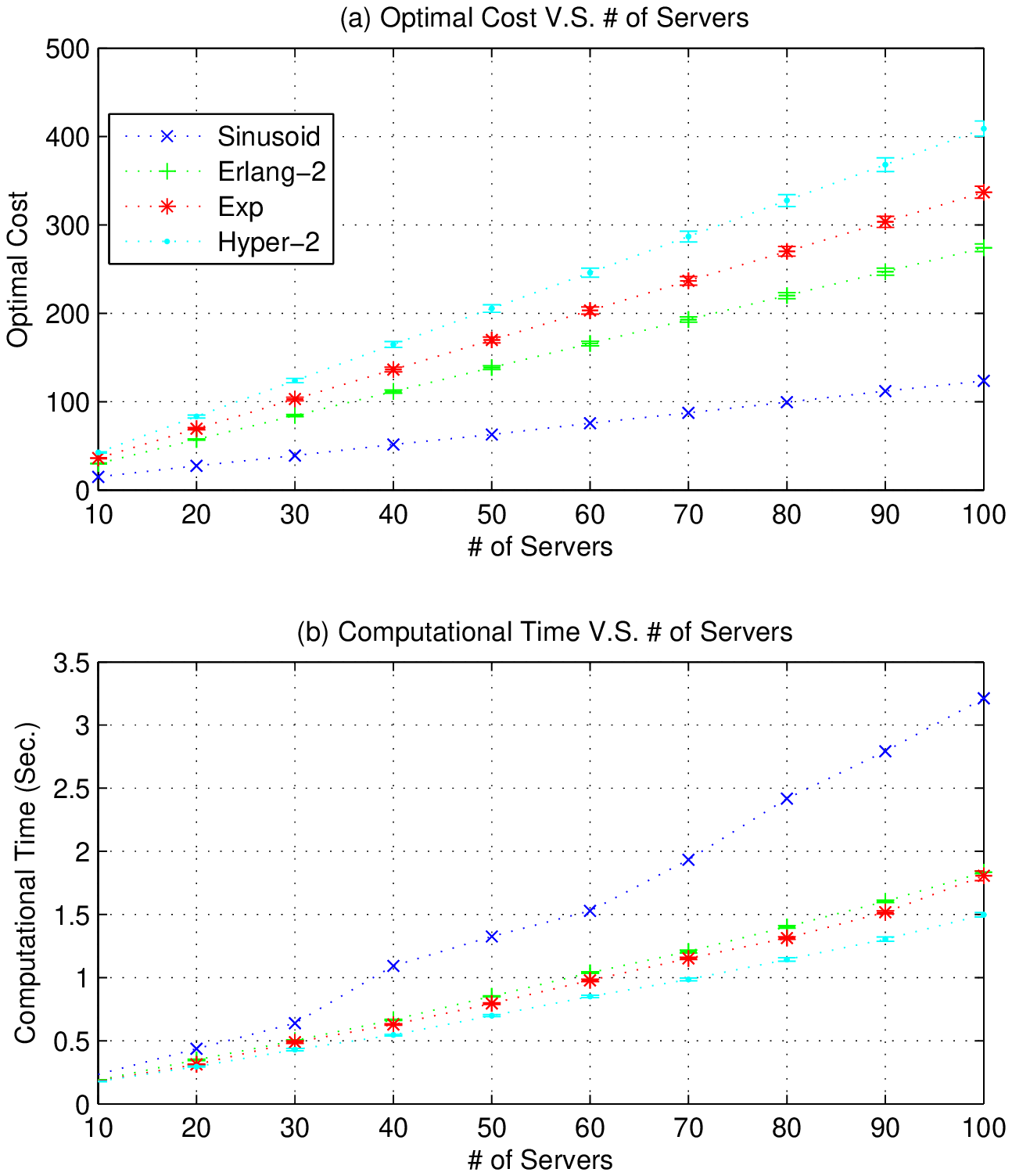}
\caption{Cost and Computational Time (with respect to Different Number of Servers in a Heterogeneous Model with 96 Time Slots)}  \label{vary_servers}
\end{minipage}
\begin{minipage}{.48\linewidth}
\centering \epsfxsize=\textwidth \epsffile[ 108   157 502
634]{./figures/Cost_Time_vary_slots.eps} \caption{The Cost
and computation time with respect to different number of slots
in a heterogeneous model with 100 servers} \label{vary_slots}
\end{minipage}
\end{figure}

We first study how the number of servers and the number of time slots of the workload affects computational time. 
We fix the number of time slots in the workload as 96 and vary the number of servers from 10 to 100 with an incremental step of 10. The optimal cost and computational time is shown in \figureref{vary_servers}(a) and (b), respectively. The optimal cost and computational time are both linear with respect to the number of servers. Then we fix the number of servers at 100 and vary the number of time slots in the workload from 10 to 100 in an incremental steps of 10 slots. The optimal cost and computation time is shown in \figureref{vary_slots}(a) and (b), respectively. Note that \figureref{vary_slots}(b) is on a log scale on the $y$-axis. In \figureref{vary_slots}(a), the optimal cost is linear with respect to the number  of slots. From \figureref{vary_slots}(b), we can see that the computational time is nearly exponentially increasing with respect to the number of slots\commentHQdelete{The pattern is between linear and exponential. The reason is that the computational time is too small to observe the pattern.}. The computational time increases linearly with respect to the increase in the number of servers \comment{while exponentially with respect to the number of time slots} because increasing the number of servers does not increase the number of constraints.
\comment{ while increasing the number of time slots does.}

\begin{figure}[!t]
\centering
\epsfxsize=.4\textwidth
\vspace{-.2in}
\epsffile[97  239  514  552]{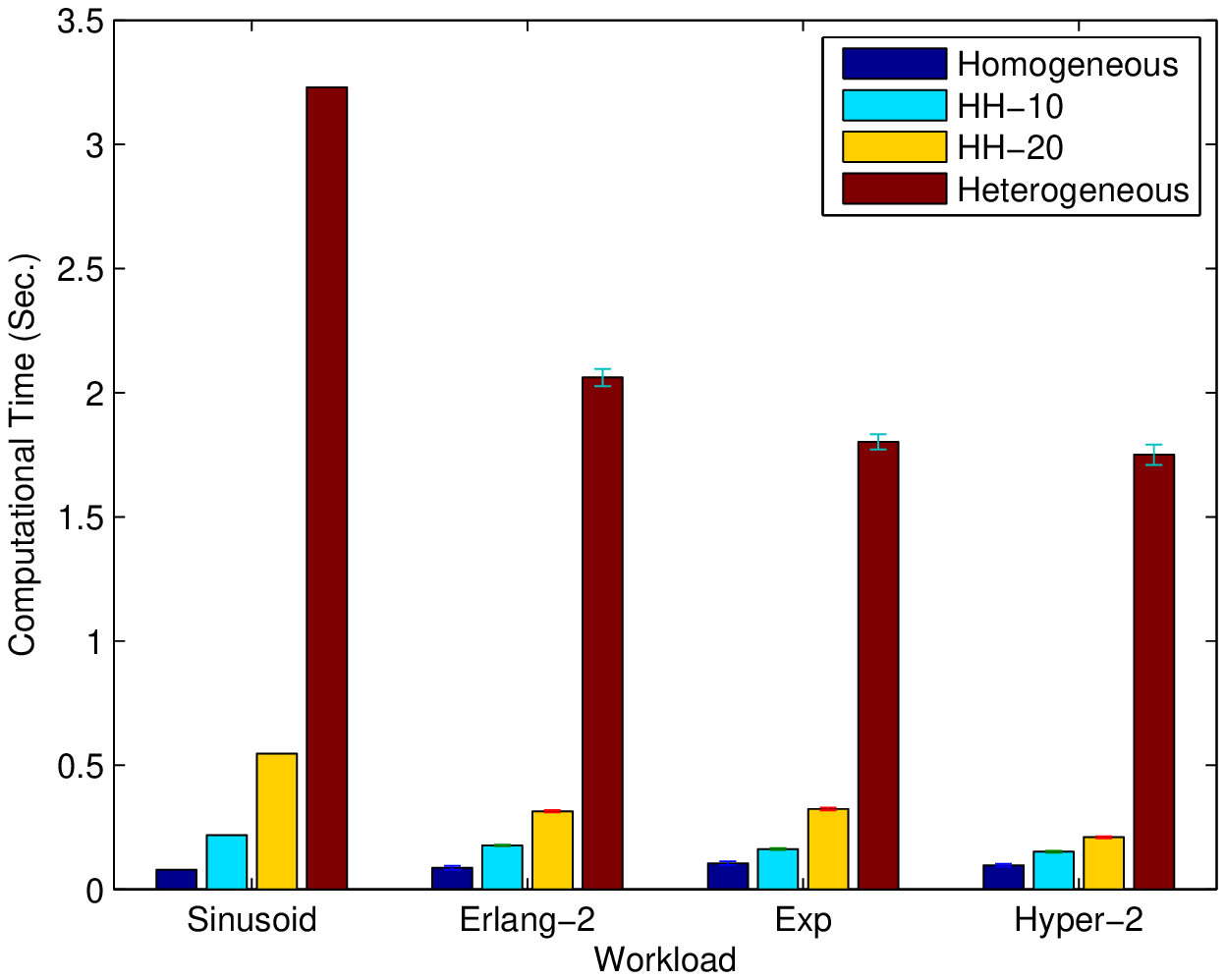}
\caption{Computational Time of Four Different Demand Types: Sinusoidal, Erlang-2, Exponential and Hyper-exponential-2 (for 100 heterogeneous Servers)} \label{homo_heter}
\end{figure}

Next, to quantify the difference in computation time among
three proposed models, we run the same problem in three models.
The data center consists of 100 identical servers (i.e.,
$I=100$). The workload with 96 time slots is generated
according to the four aforementioned distributions. This
problem fits into \textsl{Model-Hom}. To run
\textsl{Model-Het},we assume that  the servers are different
(although they are not) so that we can make comparisons.
\comment{Thus we assign a binary variable for each server at
each time slot and model this problem by Model Het.} By
dividing the servers into 10 and 20 clusters, we consider three
instances of \textsl{Model-HH}; these three instances are
denoted by HH-10 and HH-20, respectively.
\figureref{homo_heter} shows that the homogenous case has the
least computational time while the heterogeneous case has the
most computational time consistently for all four workloads.
The computational time in heterogeneous case is ten times more
than that for the homogeneous case. It also shows that a lesser
number of clusters results in less computational time for
Model-HH.

For the same workload distribution for a specific generated
seed, we obtain the same optimum; since multiple seeds are
generated for each distribution, we also present the confidence
interval on the optimal solution for each distribution, which
is summarized in Table~\ref{Tab:homo_heter_optimal}; Because
the sinusoidal workload generated is deterministic, there is no
confidence interval for the solutions of this workload. In
Table~\ref{Tab:homo_heter_optimal}, \textsl{Fixed
Configuration} means the solution obtained from statically
keeping all servers running all the time; \textsl{Local
Optimum} is the solution when the switching on and off cost is
\textit{not} considered; \textsl{Global Optimal} represents the
optimal solution obtained from our method. For the cost of
Local Optimum, we show its \textsl{Switch Cost} component. Note
that Switch Cost here includes wear-and-tear cost and the
energy cost of performing consolidation and switching on and
off. Due to the significance of Switch Cost, the result from
Local Optimum is worse than that from Fixed Configuration for
Exponential and Hyper-2. Switch Cost is correlated to the
regularity of workload shape. By applying Global Optimum, the
cost savings over Fixed Configuration and Local Optimum are
significant. Compared to the Static Configuration, we achieve
approximately 78\%, 51\%, 40\%, and 27\% savings for the
sinusoidal workload, Erlang-2, Exponential and Hyper-2,
respectively. Compared to the Local Optimum, we achieve
approximately 46\%, 46\%, and 41\% for Erlang-2, Exponential
and Hyper-2, respectively. Note that  the local optimum is
equal to the global optimum in the sinusoidal workload; this is
because of the structure of the sinusoidal workload.
\comment{Although we report results for four load
distributions,  understanding the influence of different
workload distributions on the optimal cost is outside the scope
of this paper.}
\commentHQok{To the
best of our knowledge, \cite{Qian_ITC2011} is the first effort
to decompose the workload into certain substructures and compute
the optimal solution by the substructures. A more theoretical study
is presented in \cite{Lu_SimpleAndEffective}.}
\comment{ the table with switch cost and energy cost
\begin{table*}[htbp]
\caption{Optimal Cost for Different Cases}\label{Tab:homo_heter_optimal}
\centering
\begin{tabular}[t]{c|c|c|c|c}
\hline
Workload Type&Sinusoidal&Erlang-2&Exp&Hyper-2\\
\hline
Fixed Configuration&561.1667&561.2562$\pm$0.1661&561.0864$\pm$0.1657&561.0731$\pm$0.2559\\
\hline
Local Optimum&123.5017&507.2604$\pm$8.1048&633.0925$\pm$12.8482&699.5993$\pm$19.5957\\
\hline
Switch Cost of Local Optimum&8.76&391.8281$\pm$7.1364&518.0794$\pm$11.4349&596.9251$\pm$ 17.2464\\
\hline
Energy Cost of Local Optimum&114.7417&115.4359$\pm$1.4381&115.0131$\pm$1.7569&102.6741$\pm2.5808$\\
\hline
Global Optimum&123.5017&274.2851$\pm$4.3093&337.0227$\pm$6.8391&409.0129$\pm$8.5715\\
\hline
\end{tabular}
\end{table*}
}
\begin{table*}[htbp]
\caption{Optimal Cost for Different
Cases}\label{Tab:homo_heter_optimal} \centering
\begin{tabular}[t]{c|c|c|c|c}
\hline
Workload Type&Sinusoidal&Erlang-2&Exp&Hyper-2\\
\hline
Fixed Configuration&561.1667&561.2562$\pm$0.1661&561.0864$\pm$0.1657&561.0731$\pm$0.2559\\
\hline
Local Optimum&123.5017&507.2604$\pm$8.1048&633.0925$\pm$12.8482&699.5993$\pm$19.5957\\
\hline
Switch Cost of Local Optimum&8.76&391.8281$\pm$7.1364&518.0794$\pm$11.4349&596.9251$\pm$ 17.2464\\
\hline
Global Optimum&123.5017&274.2851$\pm$4.3093&337.0227$\pm$6.8391&409.0129$\pm$8.5715\\
\hline
\end{tabular}
\end{table*}
\subsection{Insights on Aggregation }

We now present three insights that we have learned from aggregation.

\noindent \textsl{Insight-1}: We define the degree of
aggregation as the ratio of the number of original demands and
aggregated demands. Denoting the degree of aggregation by
$\alpha$, we have $\alpha = \lceil T/\hat{T}\rceil$.
Aggregation by maximum always causes over-provisioning. The
energy required to keep more than the necessary servers running
is wasteful  due to over-provisioning. On the other hand, the
aggregation by maximum smoothes out the regularity of the
workload, i.e., the fluctuation of the workload is alleviated,
which loosens the switching requirements. Moreover, the higher
the degree of aggregation,  the more smooth the workload
becomes. \textsl{Strict} dynamic aggregation is better than
static aggregation only in terms of avoiding as much
over-provisioning as possible. As to the second component of
the cost, i.e., the switching cost, it is difficult to conclude
whether dynamic aggregation is better than static aggregation
or not. Consequently, when considering the total of two cost
components, it is possible that the total cost of static
aggregation is even less than dynamic aggregation. This happens
in the case that static aggregation gains more switching cost
savings than dynamic aggregation and this difference is larger
than the gain of over-provisioning energy consumption of
dynamic aggregation over static aggregation. For the same
reason, the optimum of the high degree of aggregation may be
better than that of the low degree of aggregation.

\noindent \textsl{Insight-2}: The local smooth implementation
may \textsl{not} be better than static aggregation in terms of
avoiding as much over-provisioning as possible since the
implemented algorithm is an approximation based on local
information. However, the local smooth implementation always
favors a smoothed workload, and thus, tends to reduce switching
cost.

\comment{The strict dynamic aggregation by maximum aims at
reducing over-provisioning to overcome the pitfall of static
aggregation by maximum. As we have mentioned before, the
\textsl{strict} dynamic aggregation by maximum results in less
over-provisioning than static aggregation by maximum. Then
which one is better decided by both the absolute sum of
aggregated workload (i.e., the energy consumption component)
and the fluctuation of the workload (i.e., the switching
cost).} \noindent \textsl{Insight-3}: In aggregation by mean,
the static aggregation and dynamic aggregation end up with the
same level of average offered capacity since ``mean" is used.
That is, aggregation does not cause over-provisioning and thus,
the energy consumption of static and dynamic aggregation by
mean is equivalent. Thus, whether the aggregation method costs
less is solely decided by the switch cost, which is impacted by
the fluctuation of the workload during the planning window.

\subsection{Aggregation by Maximum\label{sec: results: aggregation by maximum}}

We now study the pros and cons of aggregating the workload by
maximum. For this study, we consider 5,000 identical servers in
a data center and the servers are clustered into fifty
100-homogenous-server groups for the purpose of management.
Thus, this falls into Model-HH. We run the optimization for
this system with different degrees of aggregation i.e., $\alpha
= 1, 2, 3, 6, 8, 12$. Note that $\alpha =1$ means that there is
no aggregation.

\begin{figure}[!t]
\centering \epsfxsize=.5\textwidth \epsffile[30   240   612
550]{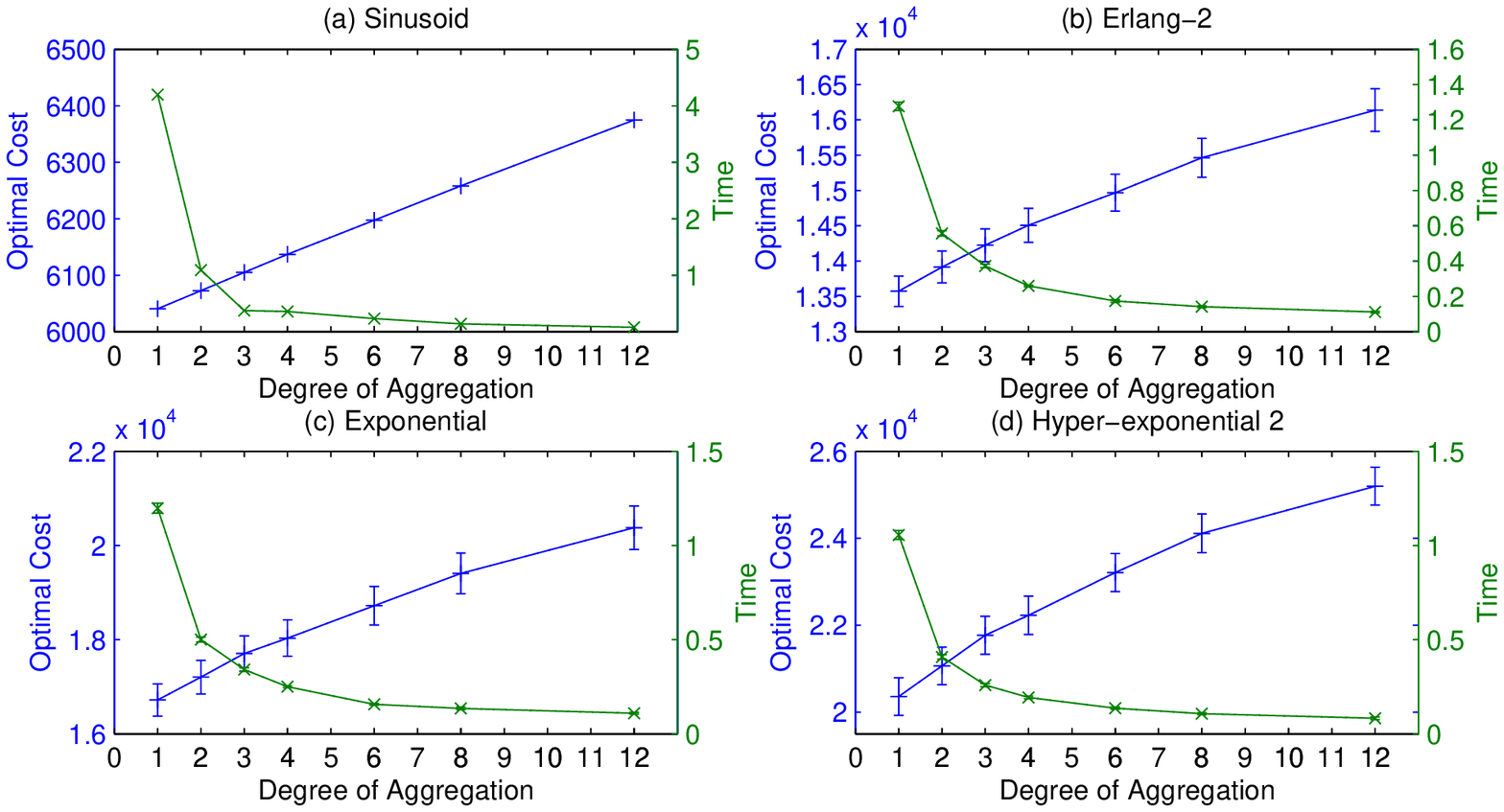}
\caption{Optimal Cost and
Computational Time with Static Aggregation by Max}
\label{static_aggregation}
\end{figure}

We first consider static aggregation. As shown in
\figureref{static_aggregation}, the cost at optimum increases
while the computational time decreases when the degree of
aggregation increases.  The gradient of the computational time
with regard to the degree of aggregation decreases as the
degree of aggregation goes up in all four cases. This means
that the degree of aggregation in a smaller range
(\textsl{e.g.}, 1-3 in this experiment) has a more significant
effect than that in a larger range (\textsl{e.g.}, 6-12 in this
experiment). Compared to the computational time for
$\alpha=12$, the computational time for $\alpha=1$ increases
more than 100 times for all four workload cases. This also
confirms that the computational time pattern increases
exponentially with respect to the number of time slots. On the
other hand, the gradient of the optimum, with regard to the
degree of aggregation, does not change  much in the entire
observed range in all workloads. Compared to the optimal cost
for $\alpha=1$, the cost at optimality for $\alpha=12$ only
increases by approximately 5\%, 16\%, 19\%, 24\% for
Sinusoidal, Erlang-2, exponential and hyper-exponential-2
cases, respectively. It is observed that a small degree of
aggregation reduces the computational time noticeably  without
significantly increasing the overall cost.
{\figureref{static_aggregation_energy} presents the energy cost
in the the static aggregation by max case. The the energy cost
increases monotonically with respect the degree of aggregation
and is the dominant components in the cost structure.}
\commentHQdelete{Computing the optimal solution by decomposing
workload is studied in \cite{Qian_ITC2011,
Lu_SimpleAndEffective}.}
\comment{It is interesting that for
the sinusoidal workload, the optimal cost for $\alpha=6$ is
even smaller than that for $\alpha=4$. This seems contradictory
to the general trend that higher degree of aggregation results
in larger optimum. It happens in the static aggregation for
deterministic workload case\footnote{In random workload case,
it might happen as well. But we consider 101 runs of different
independent seeds, the individual case cannot change the
general trend.}. We plot the workload demand before and after
aggregating in \figureref{sine_abnormal}. Although the the case
of $\alpha=4$ has less the number of time slots, its partition
is obviously better than $\alpha=6$ in term of the switch cost
(the second component of the total cost) since the fluctuation
is less. This problem has been mentioned in \textsl{Insight~1}. We will
soon see that dynamic aggregation does not have this problem.}
\begin{figure}[!t]
\centering \epsfxsize=.5\textwidth \epsffile[37   248   593   543]{./figures/max_static_energy_cost_1.eps}
\caption{Energy Cost and Wear-and-Tear Cost with Static Aggregation by Max}
\label{static_aggregation_energy}
\end{figure}

\comment{
\begin{figure}
\centering
\epsfxsize=.5\textwidth
\epsffile[97   282   514   508] {./figures/static_sine_abnormal.eps}
\caption{Static Aggregation for Sinusoidal Workload with $\alpha=4,6$}
\label{sine_abnormal}
\end{figure}
}

\begin{figure}[!t]
\centering
\epsfxsize=.5\textwidth
\epsffile[ 36   222   605   569]{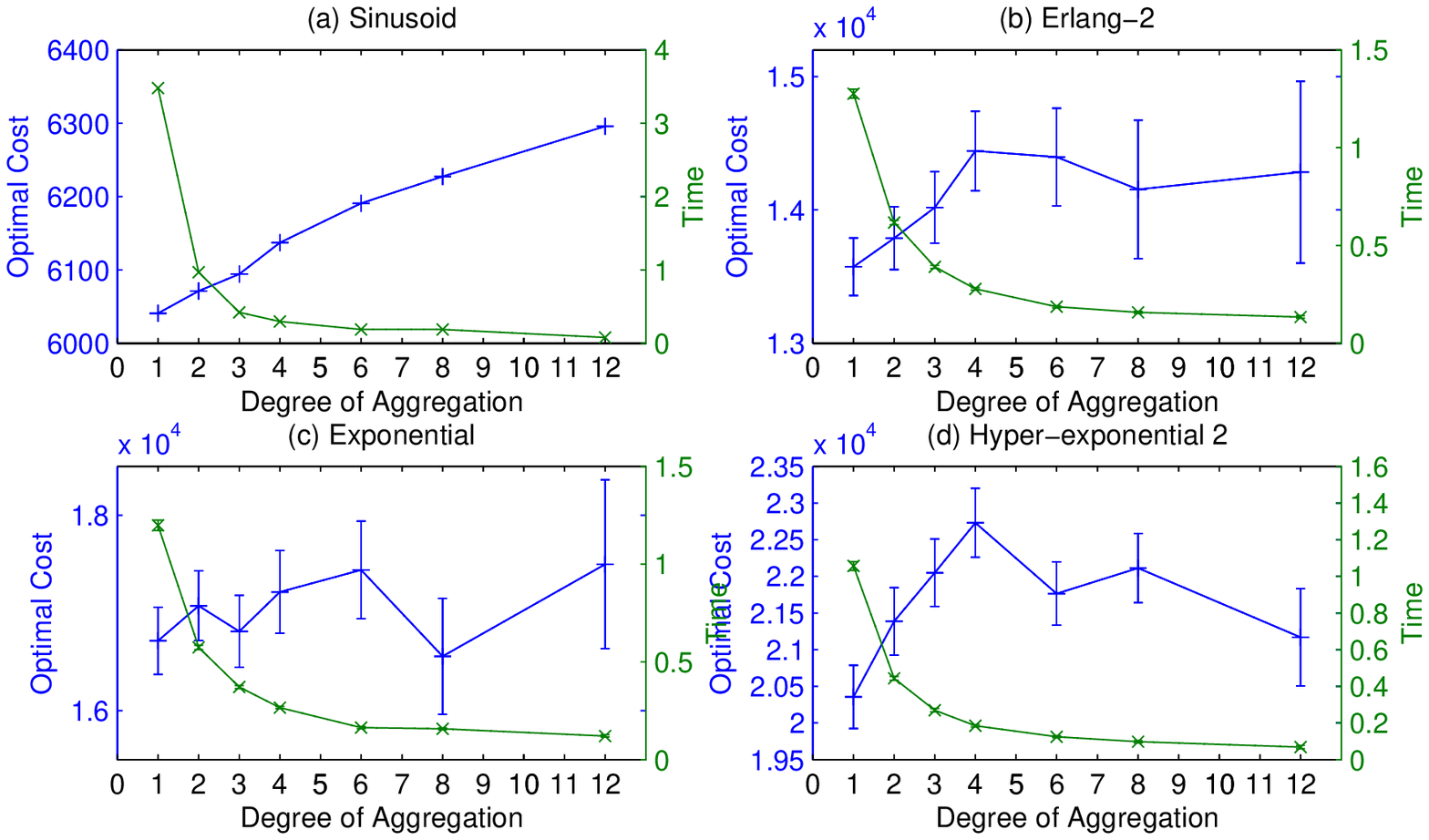}
\caption{Optimal Cost and Computational Time for Dynamic Aggregation by Maximum} \label{dynamic_aggregation}
\end{figure}

\begin{figure}
\centering
\epsfxsize=.3\textwidth
\epsffile[60   198   551   594]{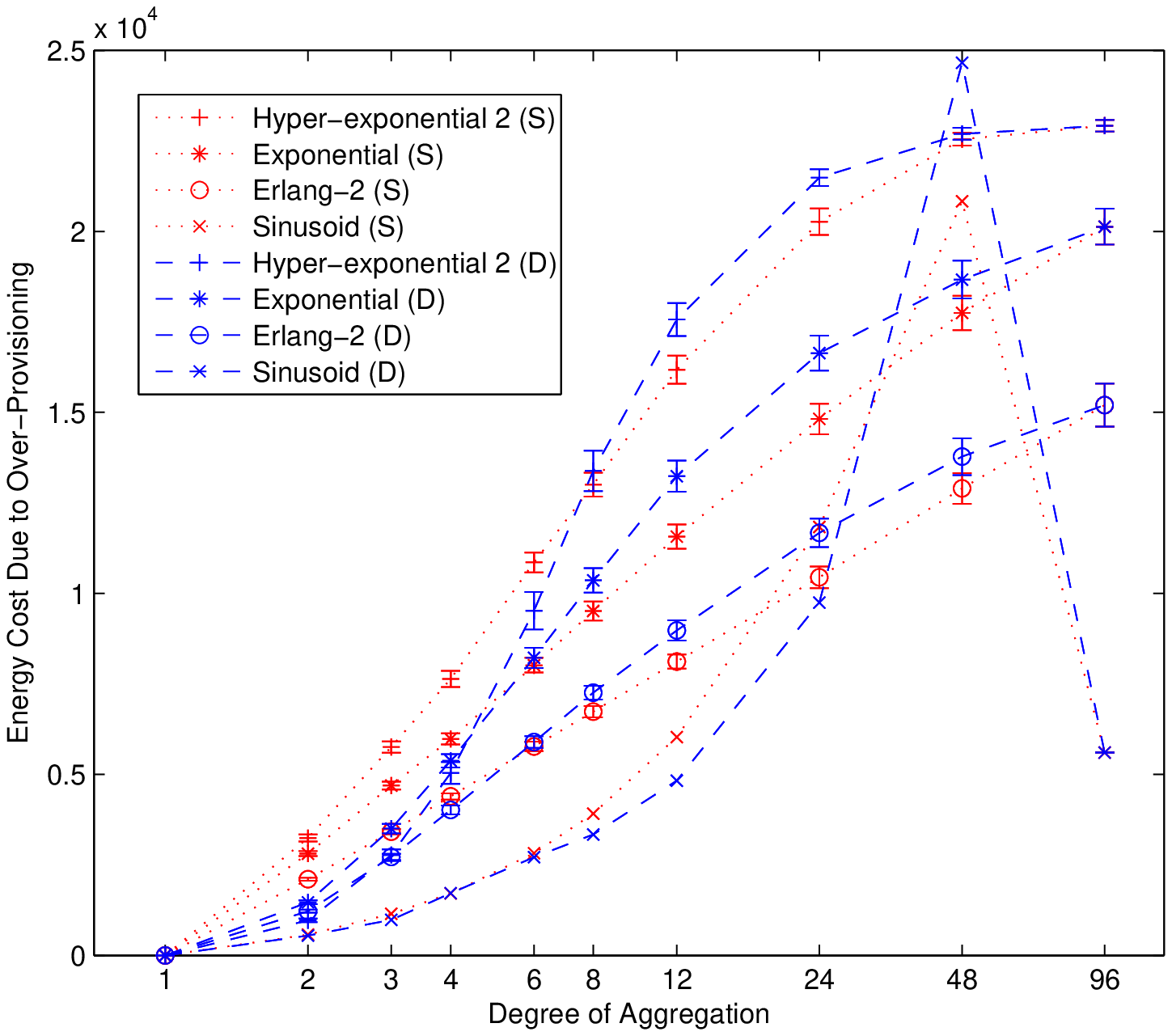}
\caption{Energy cost due to Over-provisioning: Static (S) and Dynamic (D) Aggregation by Max}
\label{max_opc}
\end{figure}

Next we consider dynamic aggregation for the same set of
workload distributions and the degree of aggregation. The
results are shown in \figureref{dynamic_aggregation}. The
pattern of change and the order of magnitude of the
computational time are similar to those in the case of static
aggregation since the number of variables and the number of
constraints in dynamic aggregation are both the same as their
counterparts in the static aggregation case. The optimum of
static aggregation consistently increases when the degree of
aggregation increases. However, this is not the case for some
workloads in dynamic aggregation. It does not always follow
this pattern in three random workload cases. It is because our
local smooth implementation of dynamic aggregation relies on
local information, which may not be able to achieve the global
optimum of minimizing energy cost but favors reducing switching
costs as we have mentioned in \textsl{Insight~1}. It is
noteworthy that (a) the violation of the pattern is relatively
minor; (b) there is no statistical difference due to the
confidence interval overlap in most pattern violation events.
More importantly, as we can see in
\figureref{static_dynamic_aggregation_optimum}, the dynamic
aggregation outperforms static aggregation in all cases except
for $\alpha=48$ in the sinusoidal workload, $\alpha=2,3,4$ in
Hyper-exponential-2 (there is no statical difference due to the
confidence interval overlap). {\figureref{static_aggregation_energy} presents the energy cost
in the the dynamic aggregation by max case. Because the energy
cost is the dominant cost components, the total cost in
\figureref{static_dynamic_aggregation_optimum} shows similar
pattern as the energy cost.}
\commentHQok{The energy
consumption as a result of over-provisioning due to aggregation by maximum
is presented in \figureref{max_opc}. For randomly distributed
workload cases, the implemented dynamic aggregation
reduces over provisioning when the degree of aggregation is small
($\leq 4$ for Erlang-2 and Exponential, $\leq 6$ for
Hyper-exponential). This also confirms that our implemented
dynamic aggregation helps reduce workload fluctuation. As we
have shown earlier, the computational time is not related to how
to aggregate. Therefore, our implementation of dynamic
aggregation shows that it is a good choice over static
aggregation in most cases. As we can see from
\figureref{static_dynamic_aggregation_optimum}(b), (c), and (d), the
discrepancy between static and dynamic aggregation forms an
elliptical shape: it starts from 0, i.e., when no aggregation needed to be
performed for $\alpha=1$. Then the discrepancy increases, then decreases and
converges back to 0 when $\alpha=96$, i.e., when the number of time slots
is 1. Thus, the aggregated workload demand  becomes the largest
workload demand of all original time slots.}

\begin{figure}[!t]
\centering \epsfxsize=.5\textwidth \epsffile[38   249   594   542]{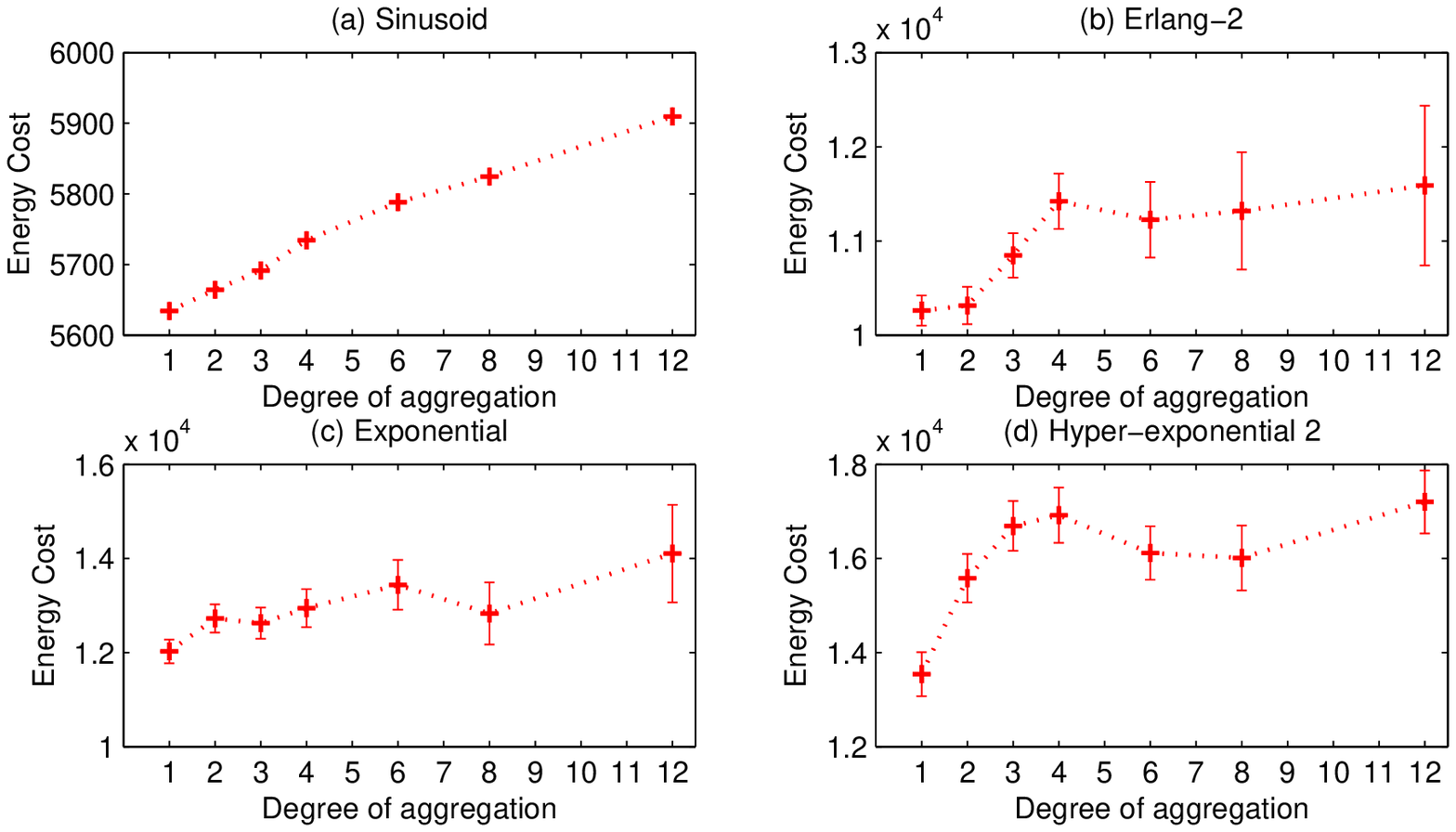}
\caption{Energy Cost and Wear-and-Tear Cost with Dynamic Aggregation by Max}
\label{dynamic_aggregation_energy}
\end{figure}

\begin{figure*}
\centering
\epsfxsize=\textwidth
\epsffile[32   180   579   611]{./figures/static_dynamic_aggregator_optimal.eps}
\caption{Comparison Between Static and Dynamic Aggregation} \label{static_dynamic_aggregation_optimum}
\end{figure*}

\subsection{Aggregation by Mean}

\begin{figure}[!t]
\centering
\epsfxsize=.5\textwidth
\epsffile[  0   239   610   552]{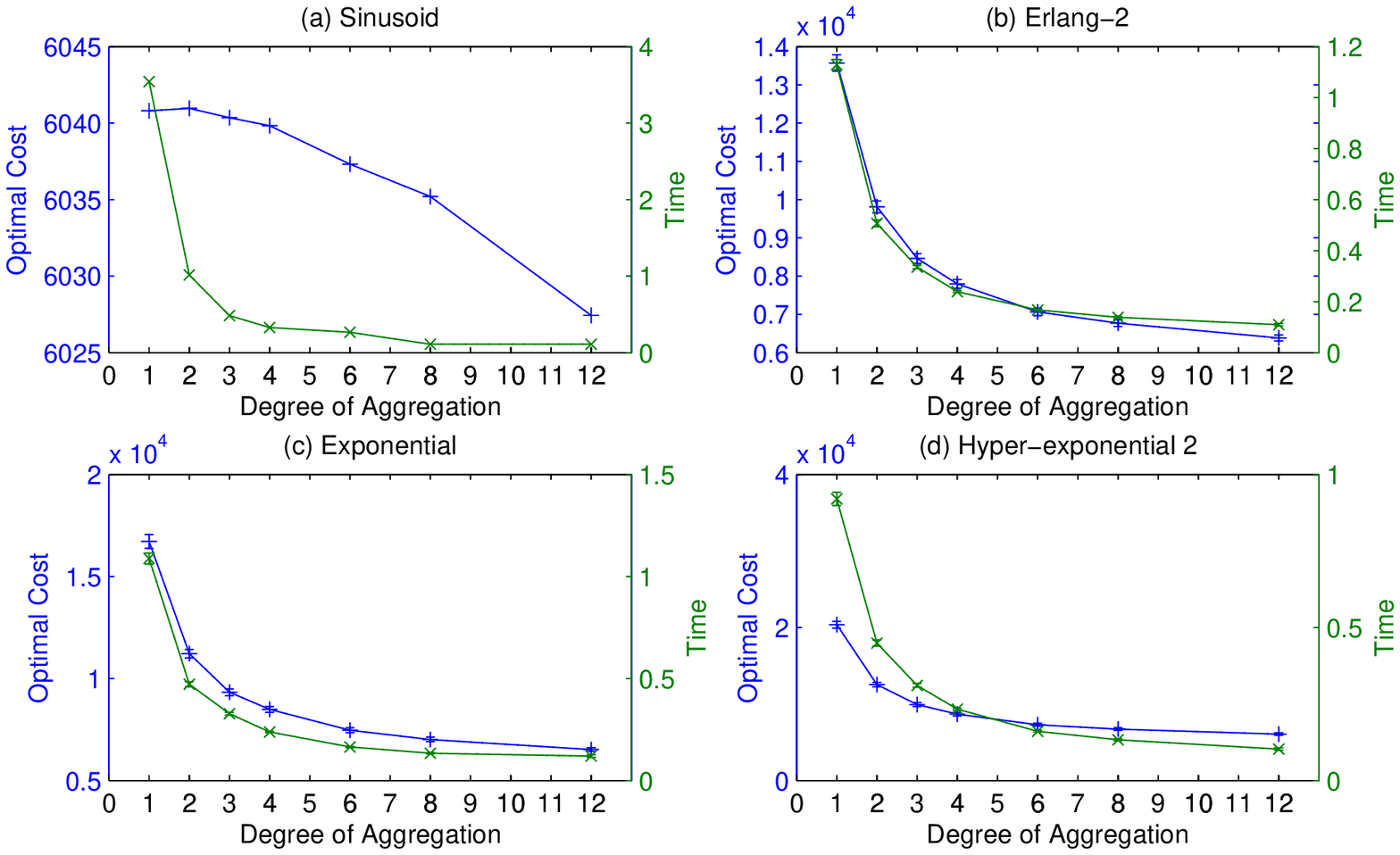}
\caption{Optimal Cost and Computational Time for Static Aggregation by Mean} \label{average_static_aggregation}
\end{figure}

\begin{figure}[!t]
\centering
\epsfxsize=.5\textwidth
\epsffile[  -50   229   664   561]{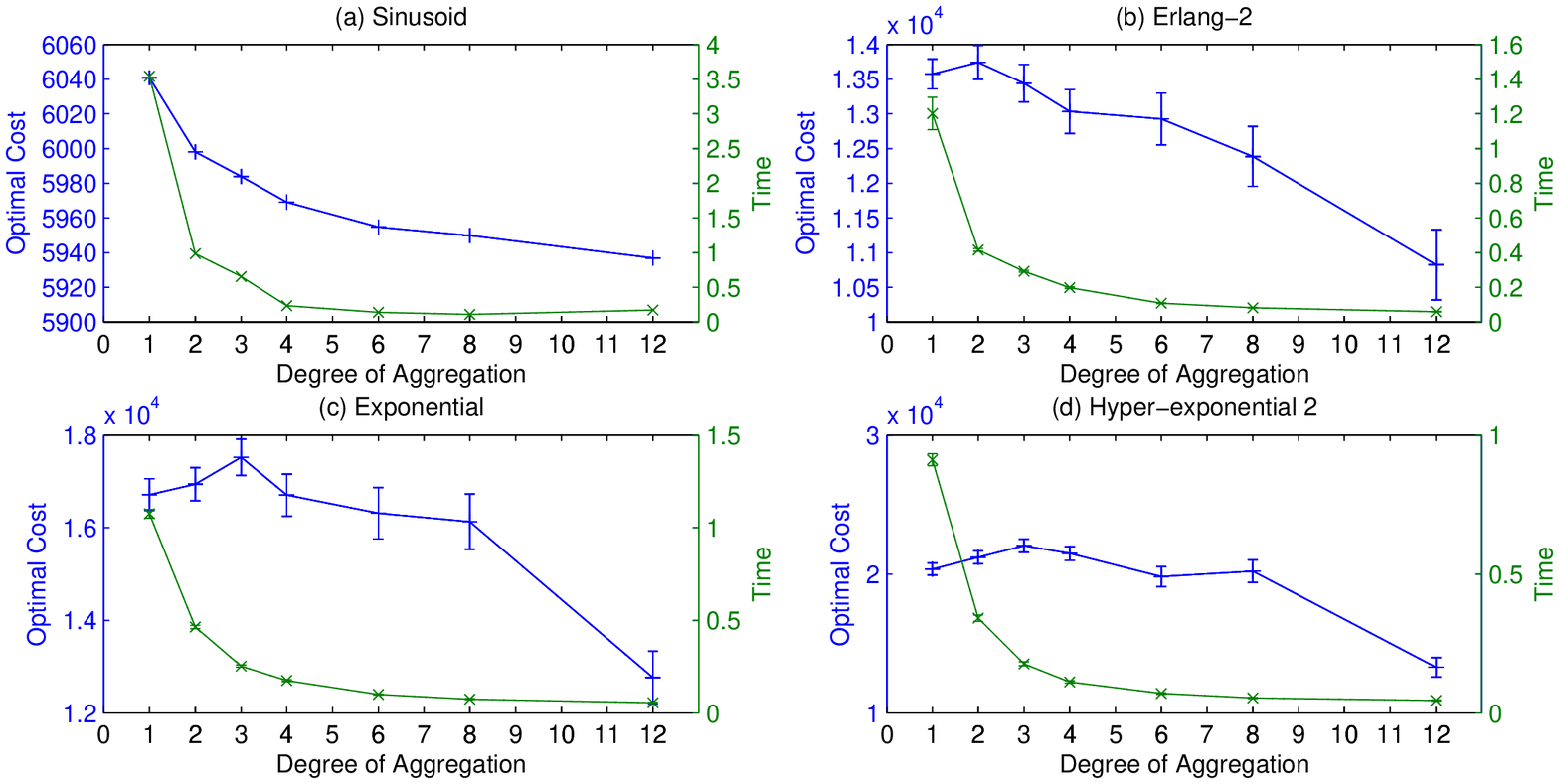}
\caption{Optimal Cost and computational time in dynamic aggregation by mean} \label{average_dynamic_aggregation}
\end{figure}

\figureref{average_static_aggregation} shows the optimal costs
and computational time when we use the static aggregation by
mean approach. In all three randomly distributed workloads, the
optimal costs and computational time are both monotonically
decreasing with respect to the degree of aggregation as we
explained in \textsl{Insight~3}. The gain for both optimum and
time complexity comes with the price of reallocating the
workload demand. The optimal costs and computational time when
applying dynamic aggregation by mean are shown in
\figureref{average_dynamic_aggregation}.

\begin{figure}
\centering
\epsfxsize=.5\textwidth
\epsffile[ -3   191   615   600]{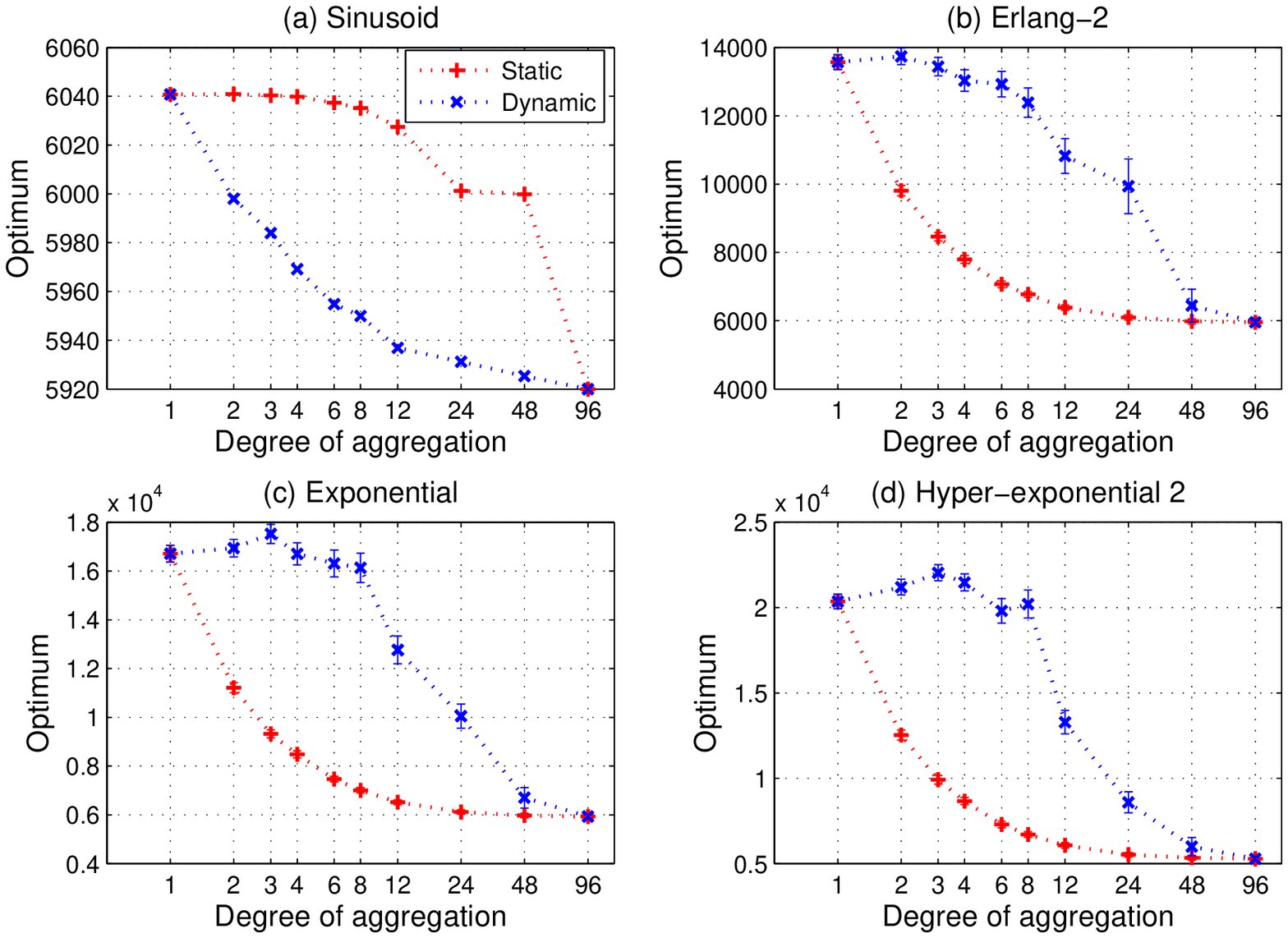}
\caption{Comparison Between Static and Dynamic Aggregation by Mean} \label{average_static_dynamic}
\end{figure}

\begin{figure}
\epsfxsize=.5\textwidth
\epsffile[-113   157   726   634]{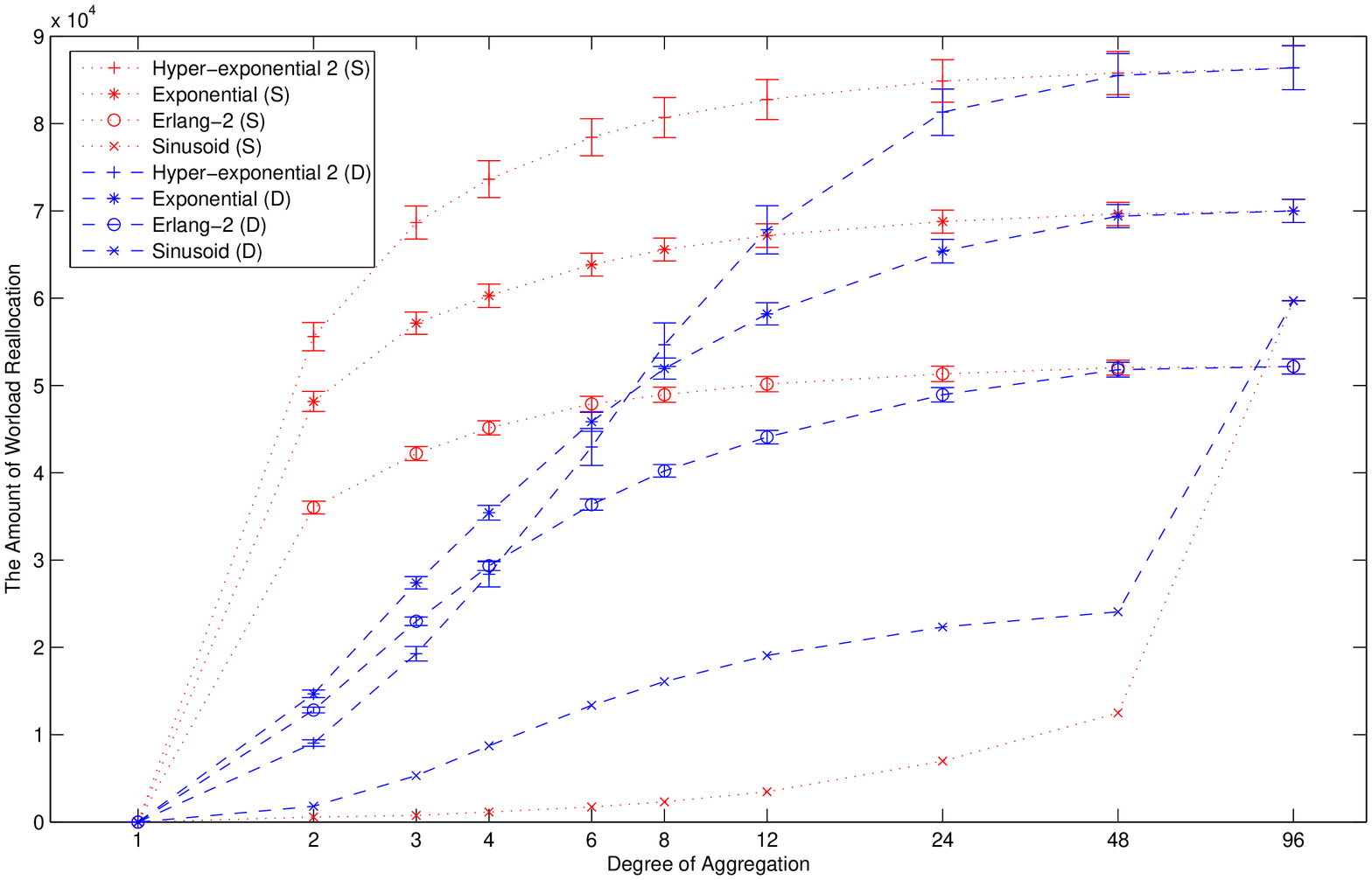}
\caption{Workload Rearrangement in Static (S) and Dynamic (D) Aggregation by Mean}
\label{workload_reallocation}
\end{figure}

Now consider comparing dynamic aggregation with static
aggregation. For computational time, there is no difference. On
the other hand, when it comes to optimal costs, conclusions are
vastly different. As shown in
\figureref{average_static_dynamic}, static aggregation
outperforms dynamic aggregation in all three non-deterministic
workload cases when aggregation by mean is considered while
dynamic aggregation outperforms static aggregation for sinusoid
workload. Recall \textsl{Insight~3}; the target of dynamic
aggregation in the aggregation by mean is to reduce
reallocating the workload. Therefore, we also present
comparative results of the amount of workload rearrangements in
both static and dynamic aggregation by mean in
\figureref{workload_reallocation}. For all three randomly
distributed workload cases, the amount of workload
rearrangements of dynamic aggregation is smaller than those of
static aggregation while that of dynamic aggregation is bigger
than that of static aggregation in the sinusoidal workload.
Thus local smooth heuristics is a good approximation for
non-deterministic workloads. This suggests that the implemented
dynamic aggregation is suitable to be used non-deterministic
workloads. As the degree of aggregation changes from 1 to 96,
the gap starts from 0, then it increases to the largest value,
then converges back to 0.

\subsection{Impacts of Varying Weights of Cost Components}

In addition, we want to understand the impact of cost
components change. This study also uses the 5,000 server
scenario that was used for the  aggregation study discussed
earlier. \commentHQok{Note that the three models yield the same
optimum but require different computational times. In other
words, there is no difference for these three models when only
optimum is considered.} We vary the weight ($\beta$) of utility
price from 0 to 1. That is, we consider the range from when the
utility price is negligible to when the wear-and-tear cost is
negligible.

\begin{figure}[htbp]
\centering \epsfxsize=.5\textwidth
\epsffile[ 0   151   600   639]{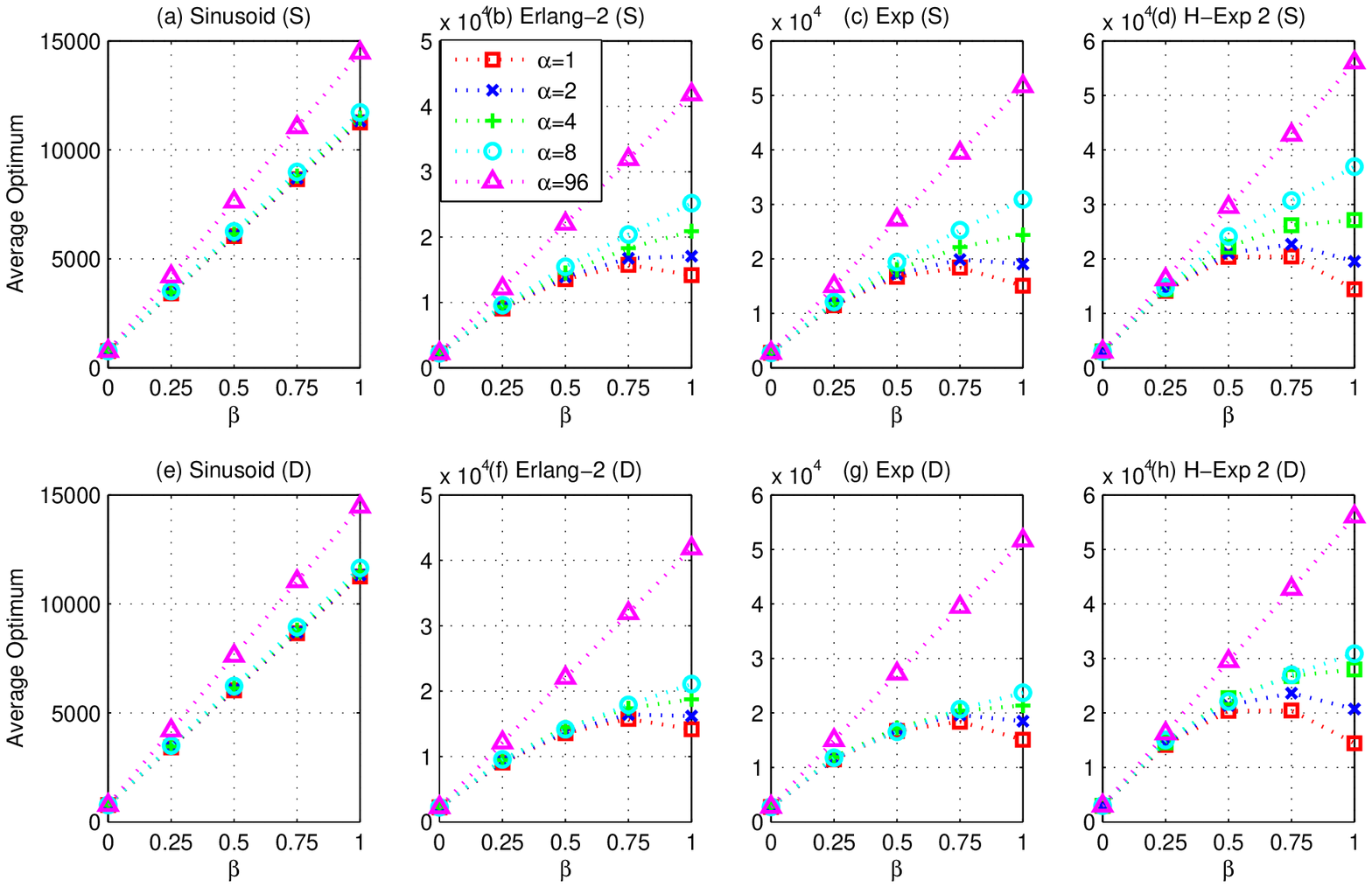}
\caption{Varying Cost Component Weight $\beta$  (5,000 Servers, Adopting Aggregation by Maximum; $\beta$ is on the $x$-axis)}
\label{change_cost_max}
\end{figure}

\figureref{change_cost_max} presents the optimum obtained under
aggregation by maximum. As we can see in
\figureref{change_cost_max}(a),(e) for sinusoidal workload, the
optimum cost is approximately linear with respect to $\beta$
for all degrees of aggregation and for both static aggregation
and dynamic aggregation. Because the switching cost in the
sinusoidal workload is very small compared to the energy cost.
This is not the case for the other three non-deterministic
workload cases since their wear-and-tear cost is comparable to
the energy cost. The exception is when $\alpha=96$, the optimum
is also linear with respect to $\beta$ since there is no
switching cost in these cases. As shown in
\figureref{change_cost_max}(b),(c),(d),(f),(g),(h), plots of
the optimum with respect to $\beta$ are concave. For $\alpha<
96$, the optimum increases first then decreases with respect to
the increase of $\beta$. It shows the dominant component to
change the optimum shifted from running energy cost
(increasing) to switching cost (decreasing). The degree of
concavity is negatively proportional to the degree of
aggregation ($\alpha$). The value of $\beta$ for optimum
``turning around" increases as the degree of aggregation
increases because the degree of aggregation goes higher, the
workload becomes smoother and causes the switching cost to be
smaller. This important observation suggests that if the weight
of switching cost is high, we can use the high degree of
aggregation to save computational time without compromising the
optimum.
\commentHQok{The gaps among
different degrees of aggregation increase as $\beta$
increases.}

In aggregation by mean, we see a similar pattern regarding the
concavity (the optimum with respect to beta) and the degree of
concavity with respect to the degree of aggregation.
\figureref{change_cost_mean} presents the same set of plots for
aggregation by mean. The optimum for static aggregation
decreases as the degree of aggregation increases because the
mean energy does not change while workload is smoothed out. The
optimum for dynamic aggregation does not have such pattern.
since the objective is not to minimize the cost.

\begin{figure}[htbp]
\centering \epsfxsize=.5\textwidth
\epsffile[ 0   151   600   639]{./change_cost_mean.eps}
\caption{Varying Cost Component Weight (5,000 Servers, Adopting Aggregation by Mean)}
\label{change_cost_mean}
\end{figure}
\comment{
\section{Application on Real Workload Trace}
}

\section{Conclusion and Future Works} \label{Conclusion}

In this paper, we first presented three formulations for
different data center environments: homogeneous data center,
heterogeneous data center, and heterogeneous
homogenous-server-cluster data center. The computational time
to obtain the optimum varies significantly in these three
cases. In order to achieve on-line (or close to on-line)
computation for large scale data centers, we proposed to
aggregate the workload to fewer time slots. Depending on the
requirements of applications and SLA allowance, there are two
types of aggregation modes. Aggregation by maximum guarantees
that the workload demand of every time slot is satisfied while
aggregation by mean needs to delay or advance the workload
demand. On the other hand, aggregation by maximum causes
over-provisioning while aggregation by mean does not.

For each of aggregation modes, we propose two aggregation methods: static and dynamic aggregation. Aggregating a fixed number of time slots into one is called static aggregation while aggregating with a certain objective is named dynamic aggregation. In the aggregation by maximum mode, the objective of dynamic aggregation is to minimize the over-provisioned capacity. In the aggregation by mean mode, the objective of dynamic aggregation is to minimize the delay and advance workload demands. An approximation implementation of dynamic aggregation is introduced to alleviate the computational overhead of implementing the exact algorithm.

Our numerical results show that aggregation is an efficient
method to reduce the computational time.  Choosing the appropriate
degree of aggregation is a tradeoff between the cost and the
computational time. We observed that the dynamic aggregating
method in both modes can achieve significant gain compared to
the static aggregation approach in terms of their individual
objective function. The sensitivity study on varying  the cost component
weights shows that the appropriate degree of aggregation also
depends on the weights.
\commentHQok{While our study is based on
artificially generated workloads using a number of random
distributions as well as a deterministic shape, the proposed
methods are general to be  applicable to realistic workloads.}
\commentHQdelete{Although we
do not consider DVFS in this paper, our aggregation methods can
be applied to the case with DVFS as well. Actually aggregation
is a more keen need in the case with DVFS since the
computational time of case with DVFS is much more compared to
the case without DVFS given other settings are the same.}

For future work, we plan to consider  decomposing and using the decomposed substructures of  a workload to determine optimal solutions. Another important direction we plan to pursue is to explore the solutions for unpredictable and partially predictable workloads. The partial predictability refers that (a) we can only accurately predict for a certain length of time, but \textsl{not} for the entire time horizon; (b) The predicted workload demand is not accurate (for example, in a certain range); (c) combination of points (a) and (b).
Results from these directions will be reported elsewhere.
\commentHQdelete{The first direction is actually the key to solve the second direction.}

\bibliographystyle{IEEEtranSdm}
\bibliography{E6Journal,Qian}

\end{document}